\documentclass[12pt,a4paper]{revtex4-1} 
\usepackage[latin1]{inputenc}
\usepackage[english]{babel}  
\usepackage{color,graphicx}   
\usepackage{amsmath}  
\usepackage{amsfonts}  
\usepackage{amssymb}   
\usepackage{mathrsfs}   
\usepackage{ulem} 
\usepackage{hyperref}
\usepackage{undertilde}
\usepackage{subfigure}
\usepackage{setspace}
\usepackage[margin=2cm]{geometry}
\usepackage[titletoc,title]{appendix}

\newcommand{\bs}[1]{\boldsymbol{#1}}
\newcommand{\mc}[1]{\mathcal{#1}}
\newcommand{\mf}[1]{\mathfrak{#1}}
\newcommand{\ul}[1]{\underline{#1}}

\newcommand{\Eq}[1]{\begin{equation}#1\end{equation}}

\hypersetup{
    colorlinks=true,
    linkcolor=blue,
    urlcolor=blue,
citecolor=blue,
}
\begin{document}
\title{Quantum heat engines: limit cycles and exceptional points}


\author{Andrea Insinga}
\email{andreainsinga@gmail.com}   
\affiliation{Department of Energy Conversion and Storage, Technical University of Denmark,\\4000 Roskilde, Denmark.}
\author{Bjarne Andresen}
\email{andresen@nbi.ku.dk}
\affiliation{Niels Bohr Institute, University of Copenhagen \\  
Universitetsparken 5, DK-2100 Copenhagen \O, Denmark}
\author{Peter Salamon}
\email{salamon@math.sdsu.edu}
\affiliation{Department of Mathematics and Statistics, San Diego State University \\
San Diego, CA 92182-7720, USA}
\author{Ronnie Kosloff}
\email{ronnie@fh.huji.ac.il}
\affiliation{Institute of Chemistry, The Hebrew University,\\Jerusalem 91904, Israel
}

\date{\today}


\begin{abstract}
We show that the inability of a quantum Otto cycle to reach a limit cycle is connected with the propagator of the cycle being non-compact.
{For a working fluid consisting of qu{a}ntum harmonic oscillators,} the transition point in parameter space where this instability occurs  is associated with a non-hermitian degeneracy {(exceptional point) of the eigenvalues} of the propagator. {In particular, a third-order exceptional point is observed at the transition from the region where the eigenvalues are complex numbers to the region where all the eigenvalues are real.} {Within this region we find another exceptional point, this time of second order, at which the trajectory becomes divergent.} {The onset of the divergent behavior corresponds to the modulus of one of the eigenvalues becoming larger than one.}
The physical origin of this {phenomenon} is that the hot and cold heat baths are unable to dissipate the frictional internal heat generated in the adiabatic strokes of the cycle.
{This behavior is contrasted with that of qu{a}ntum spins as working fluid which have a compact Hamiltonian and thus no exceptional points. All arguments are rigorously proved in terms of the systems' associated Lie algebras.}
\end{abstract}

\maketitle

\section{Introduction}
{When an engine is {started up,} typically after a short transient time it settles to a steady state operation mode: the limit cycle. An engine cycle {has reached} a limit cycle when the internal variables of the working medium {become periodic}, i.e. no energy or entropy is accumulated. Proper operation allows the engine to shuttle heat from the hot to {the }cold bath while extracting power. When the cycle time is reduced friction causes additional heat to be {generated} in the working medium. The cycle adjusts by  increasing the temperature gap between the working medium and the baths leading to increased {heat exchange. In the extreme this} leads to a situation where heat is dissipated to both the hot and cold bath{s} and  power is only consumed. But when {even }this mechanism is not sufficient to stabilize the cycle one can expect a breakdown of the limit cycle.} Here we study this phenomenon in the context of finite-time quantum thermodynamics. The working fluid of the engine {consists of} an ensemble of independent quantum harmonic oscillators.

The energy of a quantum harmonic oscillator is represented by the Hamiltonian operator $\hat{H}$, which can be written as:
\Eq{\hat{H} = \hbar \omega \Big( \hat{N} + \frac{1}{2}\Big)}
Here $\omega$ denotes the angular frequency of the oscillator and $\hat{N}$ is the number operator. The expectation value of the energy is thus determined by $\omega$ and by the expectation value of $\hat{N}$. The frequency $\omega$ is a scalar parameter which {is determined by} the dynamical laws governing the system. It can also be written as $\omega = \sqrt{k/m}$, where $k$ denotes the spring constant and $m$ denotes the mass of the oscillator. On the other hand, the number operator is related to the particular state which the system assumes: its expectation value is a measure of the degree of excitation of the system.

\begin{figure}[tb]
\centering
\subfigure{\label{fig:cycles1}
\includegraphics[width=0.48\textwidth]{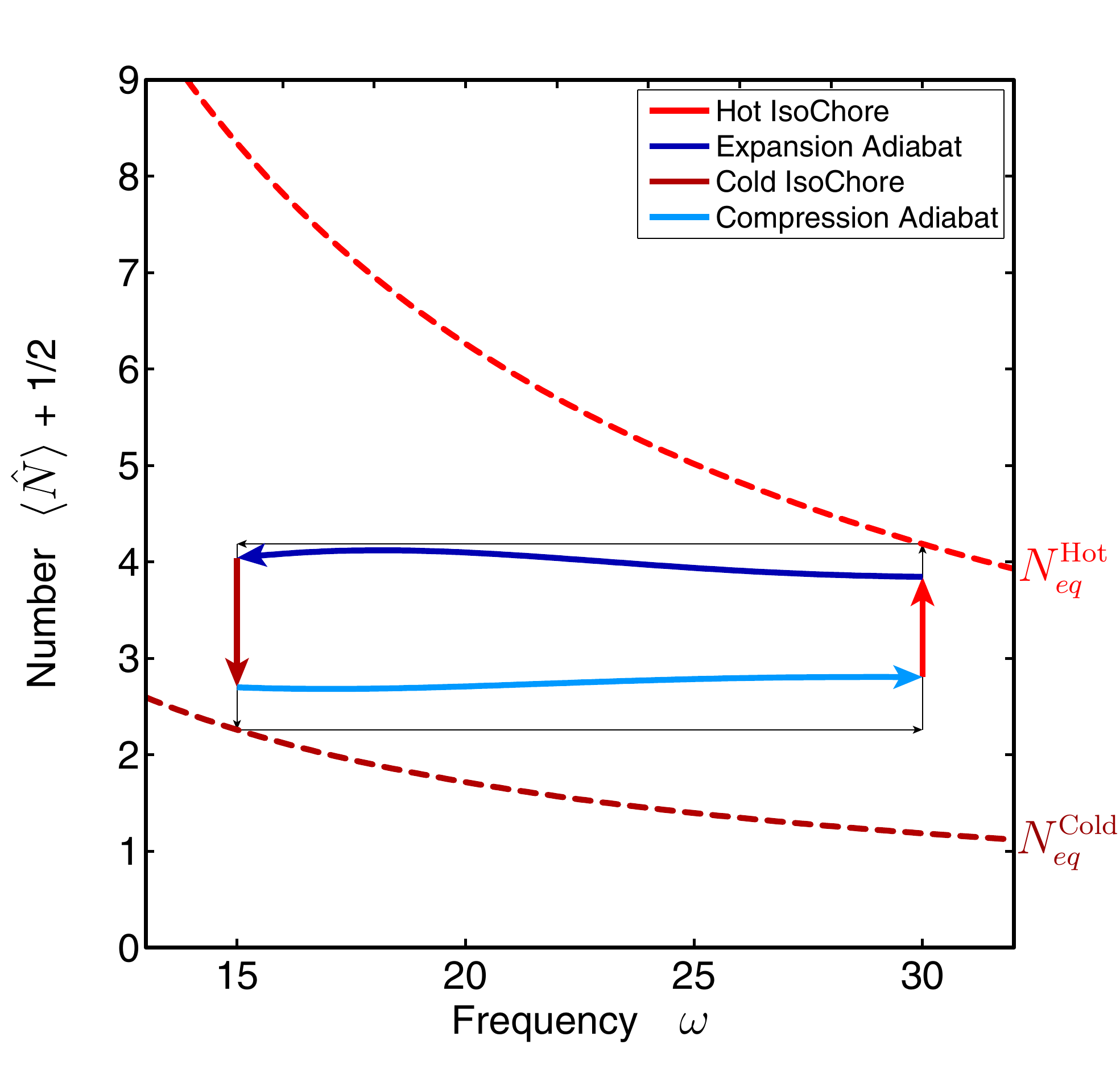}
}
\subfigure{\label{fig:cycles2}
\includegraphics[width=0.48\textwidth]{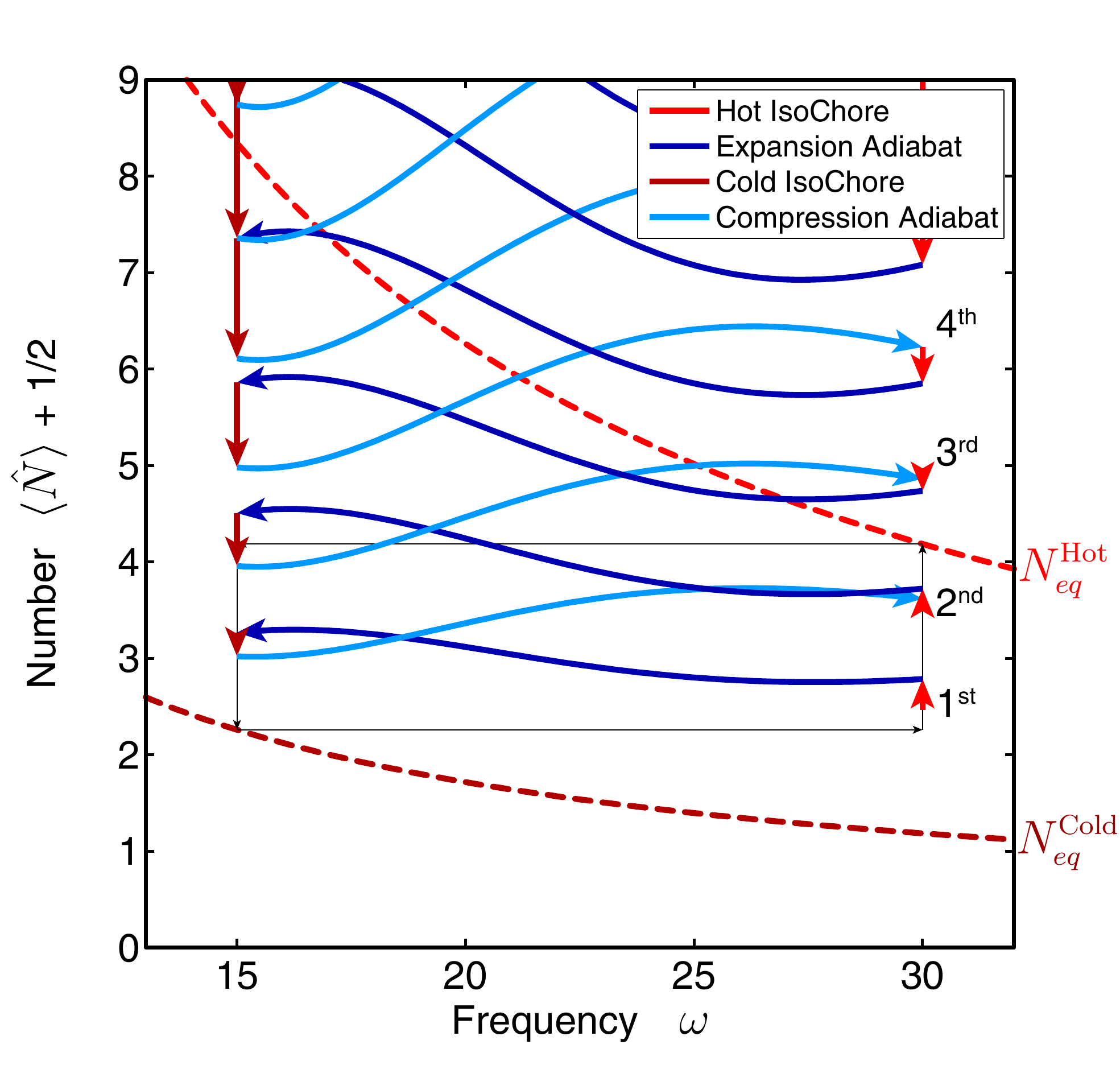}
}
\caption{Comparison between a normal cycle, in the \emph{left panel}, and a divergent cycle for which the steady state will never be reached, in the \emph{right panel}. The dashed curves represent the frequency dependence of the thermal equilibrium value of $\langle \hat{N} \rangle$ for the temperatures of the hot and cold heat reservoirs. The thin black rectangle inscribed between these curves is the long-time limit trajectory.
The times allocated for the adiabatic processes are $\tau_{HC}=\tau_{CH} =0.1$. For the left panel $\tau_H=\tau_C = 2$, while for the right panel $\tau_C = 0.4$ and $\tau_H =0.29$. The values of the other parameters are listed in Sec.~\ref{sec:OttoCycle}. } 
\end{figure}

When an ensemble of harmonic oscillators is used as working fluid of a thermodynamic machine, such as a heat engine or a refrigerator, both 
contributions to the energy {change}: the {changes} represent the energy exchange mechanisms between the working fluid and the surroundings. {Changing $\omega$ corresponds to modifying the separation between the energy levels, as happens when work is exchanged with the system, whereas changing $N$ corresponds to modifying the probability distribution among the energy levels; a change in $N$ occurs either {when heat or work is exchanged with the system.}} We can represent a thermodynamic cycle on an $(N+\frac{1}{2})$-$\omega$ diagram, reminiscent of the pressure-volume diagram which is often used to represent thermodynamic cycles of machines having a classical gas as working fluid. An example of such diagram is shown in Fig.~\ref{fig:cycles1}. This trajectory shows the quantum {analogue of} the classical Otto cycle, where the mechanical and thermal energy exchanges take place during different steps of the cycles, i.e. adiabatic and isochoric, respectively. 

As shown in Ref. \cite{Insinga2016}, and as will be discussed extensively in the present work, in the finite-time regime {there is no guarantee} that the system will converge to a limit cycle. The trajectory plotted in Fig.~\ref{fig:cycles2} shows a case where the system is not able to reach steady-state operation. The energy is poured into the working fluid cycle after cycle in the form of mechanical work, and, despite the contact with the heat reservoirs, the system is not capable of dissipating the energy fast enough. From a classical point of view this behaviour would not be surprising: nothing guarantees a priori that a system subject to a cyclic mechanical and thermal {forcing 
} will ever exhibit a periodic behaviour.

However, the Lindblad formalism, which has been introduced to describe quantum open system and the heat exchange mechanism between such systems and a thermal reservoir, has always been assumed to ensure the existence of a limit cycle solution. 
{Lindblad \cite{lindblad75} has proven that the conditional entropy decreases when applying a trace preserving completely positive map  $\mc{L}$   to both the state {represented by its density operator }$\hat{\rho} $ and the reference state $\hat{\rho} _{ref}$:
\Eq{ D( \mc{L} \hat{\rho} || \mc{L} \hat{\rho}_{ref} ) \le D(  \hat{\rho} ||  \hat{\rho}_{ref} )}
where $D(\hat{\rho} || \hat{\rho}' )= \text{Tr}\left[\hat{\rho}(\log \hat{\rho} - \log \hat{\rho}')\right]$ is the conditional entropy distance {between the states $\hat{\rho}$ and $\hat{\rho}_{ref}$}. An interpretation of {this inequality} is that a completely positive map reduces the distinguishability between two states. This observation has been employed to prove the monotonic approach to equilibrium, provided that the reference state $\hat{\rho}_{ref}$ is the only  invariant of the mapping $\mc{L}$, i.e., $\mc{L} \hat{\rho}_{ref}= \hat{\rho}_{ref}$ \cite{frigerio1977quantum,frigerio1978stationary}. The same reasoning can prove monotonic approach to the limit cycle \cite{feldmann04}. The mapping imposed by the cycle of operation of a heat engine is a product of the individual evolution steps along the branches composing the cycle of operation. Each one of these evolution steps is a completely positive map, so that the total evolution ${\cal U}_{cyc}$  that represents one cycle of operation is also a completely positive map. If a state $\hat{\rho}_{lc}$ is found that is a single invariant of ${\cal U}_{cyc}$, i.e., ${\cal U}_{cyc} \hat{\rho}_{lc}= \hat{\rho}_{lc}$, then any initial state $\hat{\rho}_{init}$ will monotonically approach the limit cycle. The largest eigenvalue of ${\cal U}_{cyc}$ with a value of $1$ is associated with the invariant limit cycle state ${\cal U}_{cycr} \hat{\rho}_{lc}=1 \hat{\rho}_{lc}$, the fixed point of ${\cal U}_{cyc}$. The other eigenvalues determine the rate of approach to the limit cycle.}

{The Lindblad-Gorini-Kossakowski-Sudarshan (L-GKS) formalism \cite{gorini1976completely, lindblad1976generators} has been applied to the study of many models of quantum heat engines, however in some cases it may be particularly important to address whether the underlying assumptions are verified or not.} {Can we guarantee a single non-degenerate eigenvalue of $1$?  {In all previously studied cases} of a reciprocating quantum heat engine a single  non-degenerate eigenvalue of $1$ was the only case found. The theorems on trace preserving completely positive maps are all based on $C^*$ algebra, which means that the dynamical algebra of the system is compact. Can the results be generalized to discrete non-compact cases such as the harmonic oscillator? Lindblad  in his study of the Brownian harmonic oscillator conjectured: {\em In the present case of a harmonic oscillator the condition that ${\cal L}$ is bounded cannot hold. We will assume this form for the generator with $\hat{H}$ and $\cal L$ unbounded as the simplest way to construct an appropriate model. }\cite{lindblad1976brownian}.  The master equation in Lindblad's form for the harmonic oscillator is well established \cite{alicki87,breuer02}, nevertheless the non-compact character of the resulting map has not been challenged. }

{In the present paper we will show a breakdown of the approach to the limit cycle. This breakdown is associated with a non-hermitian degeneracy of the cycle propagator. For special values of the cycle parameters the spectrum of the non-hermitian propagator ${\cal U}_{cyc}$ is incomplete. This is due to the coalescence of several eigenvectors, referred to as a {\emph{non-hermitian degeneracy}}. {This} difference between hermitian degeneracy and non-hermitian degeneracy is essential.  In the hermitian degeneracy, several different orthogonal eigenvectors  are associated with the same eigenvalue. In the case of non-hermitian degeneracy several eigenvectors coalesce to a single  eigenvector
\cite[Chapter 9]{moiseyev2011non}. As a result, the matrix $ {\cal U}_{cyc}$ is not diagonalizable.}

\section{Mathematical description}
\subsection{The equations of motion in the Heisenberg picture}
{In the Schr\"{o}dinger picture the mathematical description of the time evolution requires the introduction of superoperators, such as $\mc{L}$ and $\mc{U}_{cyc}$.} {A superoperator is a linear operator acting on the vector space of trace-class operators, such as the density operator $\hat{\rho}$, representing mixed states. We approach the problem within the Heisenberg picture. Instead of employing superoperators, the Heisenberg formalism involves a linear operator acting on the vector space of Hermitian operators (the observables). Both the trace-class operators and the Hermitian operators referred to above are defined over the underlying Hilbert space of pure states of the system.} 

In order to write the equations of motion in closed form, we need a finite set of Hermitian operators which is closed under the application of the commutator between any pair of operators in the set. Such set defines a Lie Algebra, which we will denote with the letter $\mf{g}$.  In particular, we consider a vector space over the field $\mathbb{R}$ of the real numbers, which is spanned by the set of anti-hermitian operators $\{i \hat{X}_j \}$, {where $\hat{X}_j$ denotes a hermitian operator and $i$ denotes the imaginary unit.}  We will use the symbol $\hat{}$ to indicate operators acting on the Hilbert space of the system. This vector space {of anti-hermitian operators}, equipped with the Lie brackets consisting of the commutator between operators, is the Lie algebra $\mf{g}$. {In fact a Lie algebra is defined as a vector space equipped with a binary operation called Lie {bracket} which must be bilinear, {alternating}, and must obey the Jacobi identity. The commutator obeys all these three properties: it is bilinear, it is {alternating}, meaning that $[\hat{X},\hat{X}] = 0$ $\forall \hat{X}\in \mf{g}$, and satisfies the Jacobi identity: $[\hat{X},[\hat{Y},\hat{Z}]]+[\hat{Z},[\hat{X},\hat{Y}]]+[\hat{Y},[\hat{Z},\hat{X}]]=0$ $\forall \hat{X},\hat{Y}, \hat{Z} \in \mf{g}$. A Lie algebra is associated to a Lie group: a continuous symmetry group which is compatible with a differential structure.}

For the basis $\{i \hat{X}_j \}$ the commutation relations can be expressed in term of the structure constant $\Gamma_{h}{}^{j}{}_{k}\in \mathbb{R}$, according to the following equation:
\Eq{[i\hat{X}_h,i\hat{X}^{j}] = \sum_k \Gamma_{h}{}^{j}{}_{k} \; i \hat{X}^k\label{eq:GammaDefinition}}
We will denote matrices with bold letters, as $\bs{A}$, and vectors with underlined  letters, as $\underline{B}$.  Upper indices, as in $\hat{X}^{j}$, indicate the  components of a column vector, while lower indices indicate the components of row vectors. We will denote by $\hat{\underline{X}}$ the vector of operators in the basis: $\hat{\underline{X}} = (\hat{X}_{1},\hat{X}_{2},\dots)^{T}$. It is convenient to introduce the set of matrices $\{\bs{A}_h\}$, whose coefficients  $a_{h}{}^{j}_{k}$ are equal to the coefficients of the structure constant:
\Eq{a_{h}{}^{j}_{k} =  \Gamma_{h}{}^{j}{}_{k} \label{eq:MatrixAfromGamma}}
The matrix $\bs{A}_h$ corresponds to the linear transformation $\text{ad}_{i\hat{X}_h}$ consisting of taking the commutator with the operator $i\hat{X}_h$. Using this notation Eq.~\ref{eq:GammaDefinition} is written as:
\Eq{\text{ad}_{i\hat{X}_h}(i \hat{\ul{X}}) \equiv [i \hat{X}_h, i\hat{\ul{X}}]=\bs{A}_h{} \; i\hat{\ul{X}}}
In order for a set of hermitian operators $\{ \hat{X}_j \}$ to be closed with respect to the equations of motion, {it is necessary that the Hamiltonian operator $\hat{H}$ be a linear} combination with real coefficients of the set $\{ \hat{X}_j \}$:
\Eq{ \hat{H}= \sum_h c^h \hat{X}_h \quad \text{with }c_h \in \mathbb{R},\; \forall \,h}
{Some Hamiltonians, e.g. an oscillator governed by an explicitely time-dependent {potential or a non-harmonic potential (e.g. containing a quartic term), cannot be expressed as a combination of elements of a finite-dimensional Lie algebra. In that case the mathematical treatment discussed in this paper} cannot be applied to such systems. However, as will be discussed in Sec. \ref{sec:HarmonicOscillator}, the Hamiltonian operator describing a quantum harmonic oscillator can be expressed as a linear combination of the elements of a finite-dimensional Lie algebra {\cite{PhysRevA.87.022116}}.}
The Heisenberg equation of motion for a hermitian operator $\hat{X}_j$ which does not depend explicitly on the time $t$ is given by:
\Eq{\frac{d}{dt}\hat{X}^{j} = \frac{i}{\hbar}[\hat{H},\hat{X}^{j}] =\frac{1}{\hbar} \sum_h c^h i [\hat{X}_h,\hat{X}^{j}] =\frac{1}{\hbar}\sum_h \sum_k c^h \Gamma_{h}{}^{j}{}_{k}\, \hat{X}^k\label{eq:HeiseqOfMotion}}
The evolution equation can be written in matrix form:
\Eq{\frac{d}{dt} \hat{\underline{X}} =\frac{1}{\hbar} \sum_h c^{h}\bs{A}_h \hat{\underline{X}} = \bs{A}\hat{\underline{X}}\label{eq:EvolutionEquation00}}
where the matrix $\bs{A}$ is defined by:
\Eq{\bs{A} = \frac{1}{\hbar} \sum_h c^{h}\bs{A}_h\; \Longleftrightarrow \; a{}^{j}_{k} =   \frac{1}{\hbar} \sum_h c^{h}  \Gamma_{h}{}^{j}{}_{k} \ \label{eq:DynamicalMatrixExpansion}}

The {transposed }matrices $\bs{A}_h{}^{T}$  {correspond} to the expansion of the adjoint representation of the algebra $\mf{g}$. If $\hat{Y} =\sum_j y^j\hat{X}_j$ and  $\hat{Z}  =\sum_k z^k\hat{X}_k = [i\hat{X}_h,\hat{Y}] $, we have: $z^k =  \sum_j \Gamma_{h}{}_{j}{}^{k}y^j$.
Since a representation of a Lie algebra is a homeomorphism, the Lie brackets of the original algebra are mapped into Lie {brackets} of its representation\cite{IntroLieAlgebras}. This means that the structure constant is the same, i.e. the commutators between two matrices $\bs{A}_h{}^{T}$ and $\bs{A}_j{}^{T}$ are given by:
\Eq{[\bs{A}_h{}^{T},\bs{A}_{j}{}^{T}] = \sum_k \Gamma_{hj}{}^{k} \bs{A}_k{}^{T}}
The set of matrices $\{ \bs{A}_h\}$ will be useful in the following sections for the purpose of highlighting the invariance properties obeyed by the equations of motion.
\subsection{The time-evolution equation}\label{sec:timeEvolEquations}
We now consider the general solution to the equation of motion expressed by Eq.~\ref{eq:EvolutionEquation00}. The solution can be formally written in terms of the time-evolution matrix $\bs{U}(t)$:
\Eq{\hat{\underline{X}}(t) =\bs{U}(t)\,\hat{\underline{X}}(0)}
The matrix $\bs{U}(t)$ satisfies the following differential equation:
\Eq{\frac{d}{dt}\bs{U}  = \bs{A}\, \bs{U} \quad \text{with} \quad \bs{U}(0) = \bs{1}}
The solution to this equation can always be written in terms of the exponential of a matrix $\bs{\Omega}$:
\Eq{\bs{U} (t) = \exp \big( \bs{\Omega} (t) \big)}
Three cases exist \cite{Sakurai}.
The simplest case is when the matrix $\bs{A}$ is time-independent. In this case $\bs{\Omega}$ is given by:
\Eq{\bs{\Omega}(t) = \ t \bs{A}  }
The second case is when $\bs{A}$ is time dependent, but satisfies the  property $[\bs{A}(t),\bs{A}(t')]=\bs{0}, \, \forall t, t'${, i.e. when $\bs{A}$ has no autocorrelation}. The solution is then given by:
\Eq{\bs{\Omega}(t)  = \int_{0}^{t'} dt'\, \bs{A}(t') }
The solution, for the general case  $[\bs{A}(t),\bs{A}(t')]\neq\bs{0}$, can be written in terms of the Magnus expansion. The matrix $ \bs{\Omega}$ is written as a sum of a series:
\Eq{\bs{\Omega} (t) = \sum_{k=1}^{\infty}  \bs{\Omega}_k (t)}
The various terms of the expansion involve nested commutators between the matrix $\bs{A}$ at different time instants:
\Eq{ \setstretch{1.5} \begin{array}{l}
\bs{\Omega}_1 (t) = \phantom{\frac{1}{1}}\int_0^t dt_1 \, \bs{A}(t_1)\\
\bs{\Omega}_2 (t) = \frac{1}{2}\int_0^t dt_1 \, \int_0^{t_1} dt_2 \, [\bs{A}(t_1),\bs{A}(t_2)]\\
\bs{\Omega}_3 (t) = \frac{1}{6}\int_0^t dt_1 \, \int_0^{t_1} dt_2 \, \int_0^{t_2} dt_3 \, \Big( [\bs{A}(t_1), [\bs{A}(t_2),\bs{A}(t_3)]] +  [\bs{A}(t_3), [\bs{A}(t_2),\bs{A}(t_1)]] \Big) \\
\dots \end{array} }

{In the next sections of the present work it will be necessary to consider the latter case for which the time-evolution equation is expressed in terms of the Magnus expansion.} We will consider the equation of motion obeyed by the expectation values of the operators in the algebra. The expectation value of an operator $\hat{X}$ will be denoted by $X$.

\subsection{Equations of motions for the harmonic oscillator}\label{sec:HarmonicOscillator}
The Hamiltonian operator $\hat{H}$ is generally written in terms of the position operator $\hat{Q}$ and the momentum operator $\hat{P}$:
\Eq{\hat{H} (t) = \frac{1}{2m}\hat{P}^2 + \frac{1}{2}m (\omega(t))^2\,\hat{Q}^2}
It is convenient to consider the following real Lie algebra of anti-hermitian time-independent operators:
\Eq{  \setstretch{1.5} \begin{array}{c c c} 
{[}i\hat{Q}^2,i\hat{D}{]}  &=&-4 \hbar \;i\hat{Q}^2\\
{[}i\hat{D},i\hat{P}^2{]} &= & -4  \hbar \;i\hat{P}^2\\
{[}i\hat{P}^2,i\hat{Q}^2{]} & = & +2i \hbar \;i \hat{D}
\end{array} \label{eq:CommOriginalAlgebra}}
Here the operator denoted by $\hat{D}$ is the position-momentum correlation operator, defined as:
\Eq{\hat{D} = \hat{Q}\hat{P} + \hat{P}\hat{Q}}
Many studies \cite{RonnieKosloff2017,rezeky06,rezek09} on quantum heat machines having as working medium an ensemble of harmonic oscillators choose a different basis for the Lie algebra, namely the set of operators $\{\hat{H},\hat{L},\hat{C}\}$. The operator denoted by $\hat{L}$ is the Lagrangian, and is given by:
\Eq{\hat{L} (t) = \frac{1}{2m}\hat{P}^2 - \frac{1}{2}m (\omega(t))^2\,\hat{Q}^2}
The operator denoted by $\hat{C}$ is proportional to the correlation operator $\hat{D}$, and is often called by the same name:
\Eq{\hat{C}(t) =  \frac{1}{2}\omega (t) \Big(  \hat{Q}\hat{P} + \hat{P}\hat{Q} \Big)}
The basis  $\{\hat{H},\hat{L},\hat{C}\}$ might be insightful from a physical point of view, and also mathematically convenient for the purpose of finding an explicit solution to the equations of motion. In the present work, however, we decided to adopt the basis $\{\hat{Q}^2,\hat{D},\hat{P}^2\}$ because, not depending explicitly on the time, it will make the mathematical derivations more transparent. It is important to point out that any result is independent of the choice of basis and could be equivalently derived with any set of linearly independent operators spanning the same space.

With our choice of basis, the set of matrices $\{\bs{A}_h\}$, defined by Eq.~\ref{eq:MatrixAfromGamma}, are given by:
\Eq{  \bs{A}_1 =\hbar \left( \begin{array}{c c c} 
0 &0&0\\
-4&0 &0\\
0 & -2 & 0
\end{array} \right) ; \,   
\bs{A}_2 = \hbar \left( \begin{array}{c c c} 
+4 &0&0\\
0&0 &0\\
0 & 0 & -4
\end{array} \right) ; \,   
\bs{A}_3 = \hbar \left( \begin{array}{c c c} 
0 &+2&0\\
0&0 &+4\\
0 & 0 & 0
\end{array} \right) \label{eq:SetA123}}
Here $\bs{A}_1$, $\bs{A}_2$, and $\bs{A}_3$ correspond to the operators $\hat{Q}^2$, $\hat{D}$, and $\hat{P}^2$, respectively. As mentioned in the previous section, the matrices $\{\bs{A}_h\}$ form a real Lie algebra:
\Eq{  \setstretch{1.5} \begin{array}{c c c} 
{[}\bs{A}_1,\bs{A}_2{]}  &=&+4 \hbar \;\bs{A}_1\\
{[}\bs{A}_2,\bs{A}_3{]} &= &+4 \hbar \;\bs{A}_3\\
{[}\bs{A}_3,\bs{A}_1 {]}& = & -2 \hbar \; \bs{A}_2
\end{array} \label{eq:AmatricesCommut}}
The reason why the commutation relations of Eq.~\ref{eq:AmatricesCommut} present a minus sign, when compared to the relations for the original algebra given by Eq.~\ref{eq:CommOriginalAlgebra}, is that the matrices $\{\bs{A}_h\}$ are the transpose of the matrices  $\{\bs{A}_h{}^T\}$ giving the adjoint representation.   

The dynamical matrix $\bs{A}$ for the basis $\{\hat{Q}^2,\hat{D},\hat{P}^2\}$ is derived from Eq.~\ref{eq:HeiseqOfMotion}\cite{RezekMSc}:
\Eq{\frac{d}{dt}\left( \begin{array}{c} Q^2 \\ D \\P^2 \end{array} \right) = 
 \left( \begin{array}{c c c} 
0&+J & 0\\
-2k&0&+2J \\
0 & -k & 0 \\
\end{array} \right)\left( \begin{array}{c} Q^2 \\ D \\ P^2\end{array} \right)} 
where $k=m \omega^2$, and $J = 1/m$. Using these symbols the Hamiltonian operator is written as:
\Eq{\hat{H} = (J/2)\hat{P}^2+ (k/2)\hat{Q}^2}
Therefore, according to Eq.~\ref{eq:DynamicalMatrixExpansion}, the matrix $\bs{A}$ can be decomposed as:
\Eq{\bs{A} =  (J/2) \bs{A}_3 +(k/2) \bs{A}_1\label{eq:AdiabaticFactorization}}
It should be stressed that all the relations presented in this section retain the same form when the coefficients $J$ and $k$ , are time-dependent. During the adiabatic processes, the frequency $\omega$ is time dependent and therefore the coefficient $k=m \omega^2$ is too.

\subsection{Equations of motion during the isochoric processes}
The evolution equation for an isochoric processes, which involves heat coupling between the system and a thermal reservoir, requires the use of the Lindblad equation. For the harmonic oscillator Lindblad's equation is expressed in the Heisenberg picture as the following equation of motion\cite{rezeky06}:
\begin{equation}
\frac{d}{dt}\hat{X}^{j} =  \frac{i}{\hbar}\left[\hat{H},\hat{X}^{j}\right] + k_{\downarrow}\left( \hat{a}^{\dagger} \hat{X}^{j} \hat{a} - \frac{1}{2}\left\{ \hat{a}^{\dagger}\hat{a},\hat{X}^{j} \right\} \right) +k_{\uparrow}\left( \hat{a} \hat{X}^{j} \hat{a}^{\dagger} - \frac{1}{2}\left\{ \hat{a}\hat{a}^{\dagger},\hat{X}^{j} \right\} \right).
\label{eq:Dissipative}
\end{equation} 
Here the operators $\hat{a}$ and $\hat{a}^{\dagger}$ are the annihilation and creation operators, respectively. They are defined in terms of $\hat{Q}$ and $\hat{P}$, according to the following equations:
\begin{equation}
\ \hat{a}= \frac{1}{\sqrt{2}} \left( \left(\frac{\sqrt{m \omega}}{\sqrt{\hbar}}\right)\hat{Q} + i \left(\frac{1}{\sqrt{m \omega \hbar}}\right)\hat{P}\right)
\label{eq:HAnnihilation}
\end{equation}
\begin{equation}
\hat{a}^{\dagger}= \frac{1}{\sqrt{2}} \left( \left(\frac{\sqrt{m \omega}}{\sqrt{\hbar}}\right)\hat{Q} - i \left(\frac{1}{\sqrt{m \omega \hbar}}\right)\hat{P}\right).
\label{eq:HCreation}
\end{equation}
The two coefficients $k_{\uparrow}$ and $k_{\downarrow}$ are known as transition rates. In order to satisfy the detailed balance condition, the ratio between the transition rates must satisfy the relation $k_{\uparrow}/k_{\downarrow} = \exp( - \beta \hbar \omega )$, where $\beta=1/k_B T$ is the inverse temperature. Eq.~\ref{eq:Dissipative} is based on the assumption that the Hamiltonian operator $\hat{H}$ does not depend explicitly on the time.

The additional term in the equation of motion requires the introduction of the identity operator $\hat{1}$. In matrix form this equation can be then expressed as \cite{RezekMSc}:
\Eq{\frac{d}{dt}\left( \begin{array}{c} Q^2 \\ D \\P^2\\1  \end{array} \right) = 
\left( \begin{array}{c c c c} 
-\Gamma&+J & 0&\frac{\Gamma}{k}H_{\text{eq}}\\
-2k&-\Gamma&+2J&0 \\
0 & -k & -\Gamma &\frac{\Gamma}{J}H_{\text{eq}}\\
0 & 0& 0 &0
\end{array} \right) \left( \begin{array}{c} Q^2 \\ D \\ P^2\\ 1 \end{array} \right)\label{eq:GeneralEvolution}} 
where $ H_{\text{eq}} = (\hbar \omega/2) \mbox{coth} (\beta \hbar \omega/2)$ is the thermal equilibrium energy corresponding to the inverse temperature $\beta$, and $\Gamma = k_{\downarrow} - k_{\uparrow}$ denotes the heat conductance.  When the identity operator is introduced, we modify the definitions of the matrices $\{\bs{A}_h\}$ expressed by Eq.~\ref{eq:SetA123} by filling with zeros the coefficients corresponding to the fourth component.

\subsection{The Otto cycle}\label{sec:OttoCycle}
As mentioned in the previous section, the Lindblad form of the equation of motion is valid as long as the Hamiltonian operator is not explicitly time dependent. For this reason we select a thermodynamic cycle where the heat transfer and mechanical work transfer never occur simultaneously, i.e. the Otto cycle. During one cycle of operation of the engine, the ensemble of oscillators undergoes  the following 4 processes in order:
\begin{itemize}
\item \emph{Hot isochore} -- The frequency of the oscillators is equal to $\omega_{H}$. The ensemble is coupled to the hot heat reservoir whose inverse temperature is denoted by $\beta_{H}$. The heat conductance is denoted by $\Gamma_{H}$.
\item \emph{Expansion adiabat} -- The mechanical work exchange is caused by the frequency varying from $\omega_{H}$ to $\omega_{C}$, while the ensemble is decoupled from the heat reservoirs.
\item \emph{Cold isochore} -- The frequency of the oscillators is equal to $\omega_{C}$. The ensemble is coupled to the cold heat reservoir whose inverse temperature is denoted by $\beta_{C}$. The heat conductance is denoted by $\Gamma_{C}$.
\item \emph{Compression adiabat} -- The frequency of the system varies from $\omega_{C}$ to $\omega_{H}$, while the ensemble is decoupled from the heat reservoirs.
\end{itemize}
The times allocated for each of these four processes are denoted respectively by $\tau_{H}$, $\tau_{HC}$, $\tau_{C}$, and $\tau_{CH}$. The total duration of a complete cycle is the sum $\tau= \tau_{H}+\tau_{HC}+\tau_{C}+\tau_{CH}$.
We denote the evolution matrices for the four branches using the same notation, i.e. $\boldsymbol{U}_{H}$,  $\boldsymbol{U}_{HC}$,  $\boldsymbol{U}_{C}$, and  $\boldsymbol{U}_{CH}$. The time-evolution matrix  $\boldsymbol{U}(\tau)$  for one cycle is the ordered product of the evolution matrices for the $4$ processes:
\Eq{ \boldsymbol{U}(\tau) = \boldsymbol{U}_{CH}\boldsymbol{U}_{C}\boldsymbol{U}_{HC}\boldsymbol{U}_{H}}
Since we focus on the case of a heat engine, the frequencies and inverse temperatures satisfy the following inequalities: $\beta_{C} > \beta_{H}$ and $\omega_{C} < \omega_{H}$.

In order to facilitate the comparison between the different results presented in this work, we fix the parameters which are used to calculate all the figures corresponding to the harmonic oscillator: 
\begin{equation}
\label{eq:TheParamSet}
\omega_H = 30, \quad \omega_{C} = 15, \quad \beta_H =0.008, \quad \beta_{C} = 0.03, \quad \Gamma_{H} = \Gamma_{C} = 0.7, \quad m = 1
\end{equation}
The calculations have been carried out using the convention that the reduced Planck constant $\hbar$ is equal to $1$. The time dependence of the frequency during the adiabatic processes is selected so that the dimensionless adiabatic parameter, $\mu=\dot{\omega} / \omega^2$ is constant. With this choice the time-evolution matrix for the adiabatic processes can be calculated analytically \cite{RonnieKosloff2017,gordon00}.

The mechanical work extracted from the working medium during each adiabatic step is the opposite of the difference between the expectation value of $\hat{H}$ at the end and the beginning of the step. For example, the work extracted during the expansion adiabat is given by:
\Eq{\mathcal{W}_{HC} = -\big( H(\tau_{H} + \tau_{HC}) - H(\tau_{H})  \Big) }
The total work $\mathcal{W}_{\text{tot}}$ extracted during one cycle is obtained as the net sum of the two contributions from the compression and expansion adiabats: $\mathcal{W}_{\text{tot}} = \mathcal{W}_{HC} + \mathcal{W}_{CH}$.  The average power $\overline{\mathcal{P}}_\text{tot}$ extracted from the system is the work divided by the duration of the cycle $\tau$:
\Eq{\overline{\mathcal{P}}_\text{tot} = \frac{\mathcal{W}_{\text{tot}}}{\tau}} 

An example of a power landscape as function of the isochore times $\tau_H$ and $\tau_C$ is shown in Fig.~\ref{fig:Pow}. The white regions correspond to divergent behavior as the trajectory shown in Fig.~\ref{fig:cycles2}, when the system is not able to converge to a limit cycle. Grey regions correspond to cycles where the heat transfer has the wrong sign for at least one of the isochoric steps. Black regions correspond to cycles providing negative work. For the regions of normal operation of the engine the color indicates the total power output $\overline{\mathcal{P}}_\text{tot}$ according to the scale shown on the right-hand side of the axes. One example of such a normal {trajectory} is shown in Fig.~\ref{fig:cycles1}. Note that the border between grey and white regions does not coincide with the boundaries of the regions with real eigenvalues (see Sec.~\ref{sec:RealTransition}).

\subsection{Calculation of the limit cycle}\label{sec:LimitCycle}
We will now briefly review the procedure discussed in Ref. \cite{Insinga2016}, which concerns the determination of limit cycles, and the classification of their stability.
Since the identity operator $\hat{1}$ does not evolve with time, it is insightful to consider the analogy with homogeneous coordinate systems. From this point on we will denote with the symbol  $\tilde{}$ the $3 \times 1$ vectors and $3 \times 3$ matrix blocks acting on the first three variables $Q^2$, $D$, and $P^2$.  In this notation, the matrix $\bs{A}$ giving the equations of motion is written as:
\begin{equation}
\boldsymbol{A}(t) = 
  \left( \begin{array}{c c c |c} 
 & & & \\
 &\tilde{\boldsymbol{A}}(t)& &\tilde{\underline{B}}(t) \\
 & & &  \\
 \hline 0 &0  &0  &0 
\end{array} \right)
\end{equation}
Because of the properties discussed in Sec.~\ref{sec:PropertyBlocks}, the time-evolution equations presented in Sec.~\ref{sec:timeEvolEquations} applied to a matrix $\bs{A}$ of this form always produce a time-evolution matrix $\boldsymbol{U}$ with the following structure:
\begin{equation}
\boldsymbol{U}(t) = 
 \left( \begin{array}{c c c |c} 
 & & & \\
 &\tilde{\boldsymbol{U}}(t)& &\tilde{\underline{C}}(t) \\
 & & &  \\
 \hline 0 &0  &0  &1 
 \end{array} \right).
\label{eq:TheUMatrix}
\end{equation}
The $3 \times 3$ matrix $\tilde{\boldsymbol{U}}$ is the linear part of the evolution, the vector $\tilde{\underline{C}}$ acts as a translation in the space of the first $3$ variables.

The relation $\underline{X}(t+\tau) =\boldsymbol{U}(\tau) \underline{X}(t)$ is thus analogous to the following equation:
\begin{equation}
\tilde{\underline{X}}(t+\tau) = \tilde{\boldsymbol{U}}(\tau)\tilde{\underline{X}}(t)  + \tilde{\underline{C}}(\tau)\label{eq:TranslCycle}
\end{equation}
A point $ \tilde{\underline{X}}^{0}$ is invariant under the previous equation if at the time $t=0$ it satisfies:
\begin{equation}
\tilde{\underline{X}}^{0}= \tilde{\boldsymbol{U}}(\tau)\tilde{\underline{X}}^{0}  + \tilde{\underline{C}}(\tau) = (\tilde{\boldsymbol{1}} - \tilde{\boldsymbol{U}}(\tau))^{-1}\, \tilde{\underline{C}}(\tau) \label{eq:invariantPointX0}
\end{equation}
This equation expresses the fact that the invertibility of $\tilde{\boldsymbol{1}} - \tilde{\boldsymbol{U}}(\tau)$ is a sufficient condition for the existence of an invariant point $\tilde{\underline{X}}^{0}$, which can also be called a stationary solution. 

As is pointed out in Ref. \cite{RezekMSc}, the invertibility of $\tilde{\boldsymbol{1}} - \tilde{\boldsymbol{U}}(\tau)$ does not guarantee that the stationary solution is stable, i.e. an attractive equilibrium point. An equilibrium point $\tilde{\underline{X}}^{0}$  is attractive if it is obtained from an arbitrary initial state $\tilde{\underline{X}}(0)$ by iteratively applying the one-cycle evolution for an infinite number of cycles:
\begin{equation}
\label{eq:attractive}
\lim_{n\rightarrow +\infty}\tilde{\underline{X}}(n\tau)  = \tilde{\underline{X}}^{0}.
\end{equation}
Applying Eq.~\ref{eq:TranslCycle} of evolution for $n$ cycles can be expressed as the following factorization:
\begin{equation}
\tilde{\underline{X}}(n\tau) = \tilde{\boldsymbol{U}}^{n}(\tau) \tilde{\underline{X}}(0) + \sum_{k=0}^{n-1}\tilde{\boldsymbol{U}}^{k}(\tau)  \tilde{\underline{C}}(\tau).
\end{equation}
The first term of the right-hand side explicitly depends on the initial state $\tilde{\underline{X}}(0)$. However, the equilibrium solution can be independent of the initial state only if this term vanishes, which leads to the following requirement:
\begin{equation}
\lim_{n\rightarrow +\infty} \tilde{\boldsymbol{U}}^{n}(\tau) =  \tilde{\boldsymbol{0}}.
\end{equation}

This condition can be verified if and only if the moduli of all the eigenvalues of the matrix $\tilde{\boldsymbol{U}}(\tau)$ are strictly smaller than $1$. In this case the geometric series generated by $ \tilde{\boldsymbol{U}}(\tau)$ is convergent and its limit is given by:
\begin{equation}
\lim_{n\rightarrow +\infty}  \sum_{k=0}^{n-1}\tilde{\boldsymbol{U}}^{k}(\tau) =(\tilde{\boldsymbol{1}} - \tilde{\boldsymbol{U}}(\tau))^{-1}.
\label{eq:geomseries}
\end{equation}

\begin{figure}[httb!]
\centering

\subfigure{\label{fig:Pow}
\includegraphics[width=0.48\textwidth]{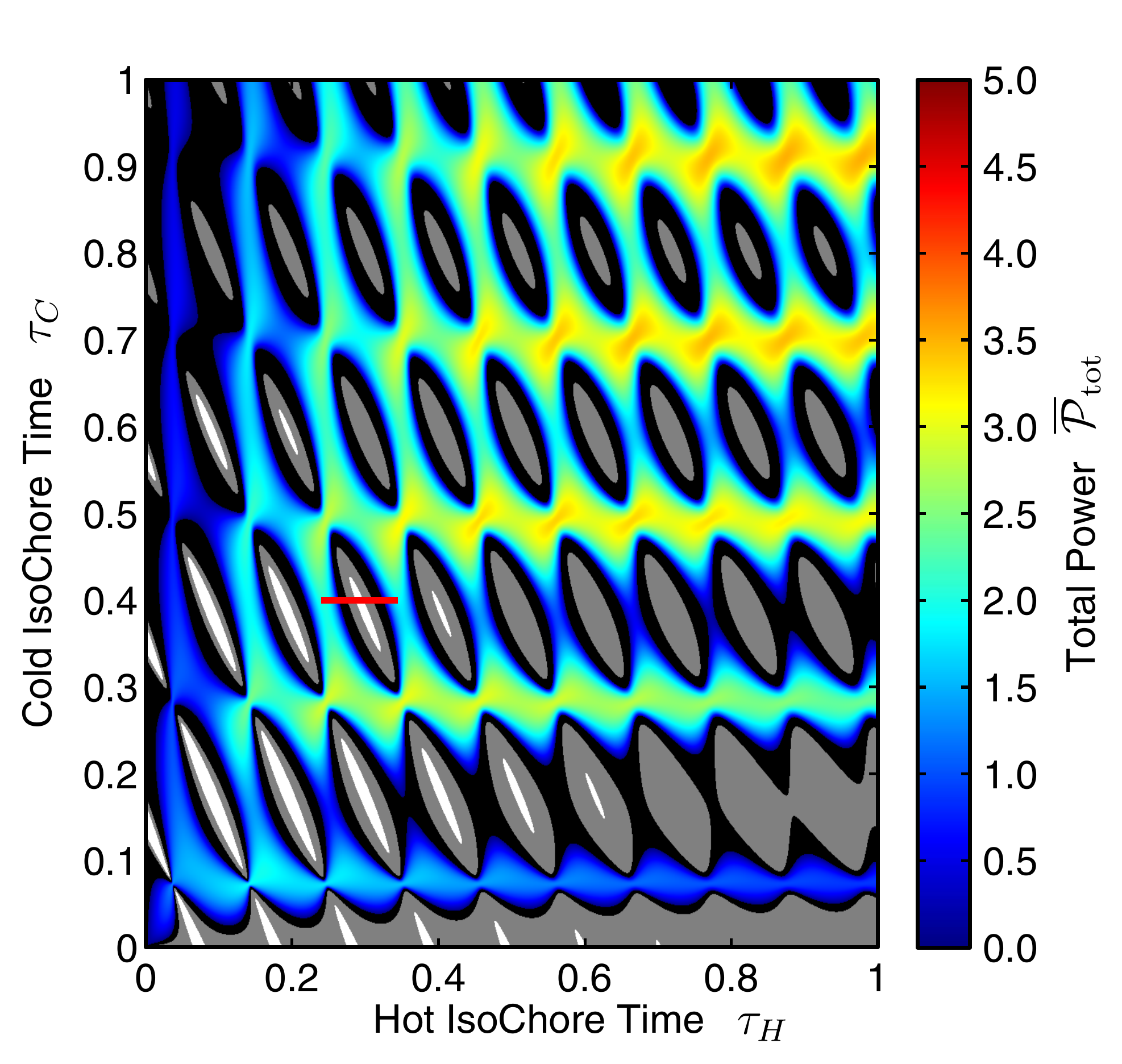}
}
\subfigure{\label{fig:eigs1}
\includegraphics[width=0.48\textwidth]{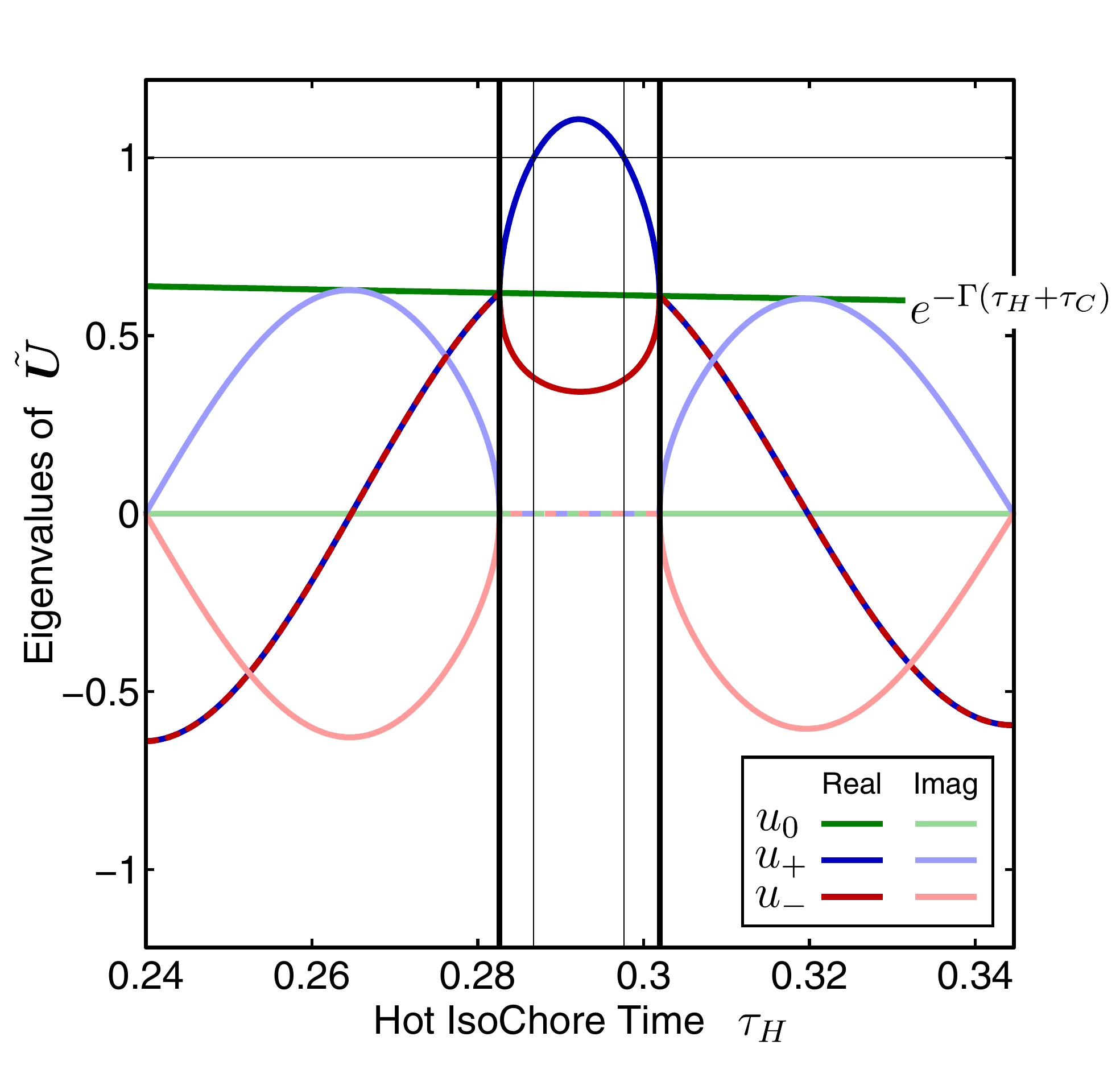}
}

\caption{\emph{Left panel}: power landscape for $\tau_{HC} = \tau_{CH} = 0.1$. The white regions correspond to choices of parameters for which the system is not able to converge to a limit cycle, as for the trajectory shown in Fig.~\ref{fig:cycles2}. \emph{Right panel}: eigenvalues of the $3\times 3$ block $\tilde{\bs{U}}$ of the time-evolution matrix for one cycle. The dark  and bright shades of the same hue indicate the real and imaginary parts of the same eigenvalues, respectively. Dashed lines indicate that the curves of the corresponding colors overlap.  When one of the eigenvalues, in this case $u_{+}$, has modulus greater than $1$, the limit cycle can never be reached. This figure corresponds to the segment highlighted by the horizontal red line shown in the left panel, $\tau_C = 0.4$.} 
\end{figure}

Therefore, when the condition is satisfied the invariant point $\tilde{\underline{X}}^{0}$ defined in Eq.~\ref{eq:invariantPointX0} is also stable. The eigenvalues of $\tilde{\bs{U}}$ are plotted in Fig.~\ref{fig:eigs1} as functions of the hot isochore time $\tau_H$.  The colours red, green and blue identify the three different eigenvalues. For each of the three colours there is a darker shade, indicating the real part, and a brighter shade, indicating the imaginary part.
As can be noticed,  in the middle region of the graph, delimited by the thick vertical black lines, all three eigenvalues are real. As we will show in the next sections, when the eigenvalues are not purely real, they are necessarily complex numbers with norm equal to  $e^{-\Gamma ( \tau_H + \tau_C)}$.

We can also see from the figure that in the smaller central region delimited by the thin vertical black lines, the eigenvalue $u_{+}$ corresponding to the blue curve is greater than $1$. In this region the system is not able to converge to a limit  cycle, behaving as the example shown in  Fig.~\ref{fig:cycles2}.

\section{The role of exceptional points}
\subsection{Decomposing the equations of motion}
In this section we consider a decomposition of the equations of motions which clarifies that the effect of the diagonal terms in Eq.~\ref{eq:GeneralEvolution} can be factored out and resolved from the remaining terms of the equations.  This factorization will be used in the next section to highlight the nature of the transition between the oscillatory behavior, when the eigenvalues of $\tilde{\boldsymbol{U}}(\tau)$ are complex, and the exponential behavior, when the eigenvalues of $\tilde{\boldsymbol{U}}(\tau)$ are real. 

We start by considering the matrices defined in equation~\ref{eq:SetA123}, which, according to the notation introduced in Sec.~\ref{sec:LimitCycle}, will be 
denoted by $\tilde{\bs{A}}_h$ since they are $3 \times 3$ matrix blocks.
Moreover, we introduce the matrix $\tilde{\bs{A}}_0$ which commutes with the other three matrices. 
\Eq{\tilde{\bs{A}}_0 = \left( \begin{array}{c c c c} 
+1 &0&0\\
0&+1 &0\\
0 & 0 & +1
\end{array} \right) \label{eq:A0martix}}
We notice that the first $3\times 3$ block of $\bs{A}$ from Eq.~\ref{eq:GeneralEvolution} can be written as:
\Eq{\tilde{\bs{A}} = -\Gamma \tilde{\bs{A}}_0 + (J/2) \tilde{\bs{A}}_3 + (k/2) \tilde{\bs{A}}_1}
This equation generalizes equation~\ref{eq:AdiabaticFactorization} by including the diagonal terms proportional to  the heat conductance $\Gamma$.
During the isochoric processes $\bs{A}$ is time independent and $\bs{U}$ can be calculated by taking the exponential of $t\bs{A}$. We now use the property expressed by Eq.~\ref{eq:ExpBlock} { in Appendix \ref{sec:PropertyBlocks}}. For the hot isochore process (and similarly for the cold one) we have: 
\Eq{ \tilde{\boldsymbol{U}}_{H} =  e^{-\Gamma  \tau_{H} } \exp \Big( \tau_{H} \big( (J/2) \tilde{\bs{A}}_3 + (k/2) \tilde{\bs{A}}_1 \big)  \Big) }
Since $ \tilde{\bs{A}}_0$ is proportional to the identity matrix $\tilde{\bs{1}}$, the exponential of the matrix $\Gamma \tilde{\bs{A}}_0$ can be written as a multiplying scalar. 
Because of the property expressed by Eq.~\ref{eq:ProductBlocks} from Sec.~\ref{sec:PropertyBlocks}, the  $3 \times 3$ block of the one-cycle evolution matrix $\bs{U}(\tau)$ can be obtained by multiplying the  $3 \times 3$ blocks of the $4$ evolution matrices corresponding to the adiabatic and isochoric processes composing the cycle:
\Eq{\tilde{\boldsymbol{U}}(\tau) = \tilde{\boldsymbol{U}}_{CH}\tilde{\boldsymbol{U}}_{C}\tilde{\boldsymbol{U}}_{HC}\tilde{\boldsymbol{U}}_{H}}
The effect of the dissipative processes on the $3 \times 3$ block $\tilde{\boldsymbol{U}}(\tau) $ is to introduce a multiplicative scalar factor $e^{-\Gamma  ( \tau_H +\tau_C)}$. 

\subsection{Transition to real eigenvalues:\\$3^{\text{\scriptsize rd}}$ order non-hermitian degeneracy}
\label{sec:RealTransition}
We will now show that \emph{the transition between real and complex eigenvalues involves an exceptional point where the three eigenvectors coalesce.}  This transition corresponds  to, e.g., the values of $\tau_H$ indicated by the thick vertical black lines of Fig.~\ref{fig:eigs1}. When the norm of the eigenvalues is smaller than one, complex eigenvalues correspond to a stable spiral, while real eigenvalues correspond to a stable node.

For now we consider the $3 \times 3$ matrix $\tilde{\boldsymbol{U}}(\tau) $ disregarding the factor $e^{-\Gamma  ( \tau_H +\tau_C)}$. Disregarding this factor is equivalent to setting $\Gamma=0$. The problem is reduced to finding the evolution matrix having time derivative given by:
\Eq{\tilde{\bs{A}}(t) =(J(t)/2) \tilde{\bs{A}}_3 + (k(t)/2) \tilde{\bs{A}}_1} 
The solution of the corresponding differential equation requires the use of a Magnus expansion {because $\tilde{\bs{A}}$ is time-dependent and exhibits autocorrelation}. It follows from the expression of the various terms appearing in the  expansion, that if $\tilde{\bs{A}}$ belongs to a Lie algebra, then $\tilde{\bs{\Omega}}$ does too, and it is always possible to express it as a linear combination of the matrices $\tilde{\bs{A}}_1$, $\tilde{\bs{A}}_2$ and $\tilde{\bs{A}}_3$:
\Eq{\tilde{\bs{\Omega}}(\tau)  =\alpha_1  \tilde{\bs{A}}_1 +  \alpha_2  \tilde{\bs{A}}_2 + \alpha_3  \tilde{\bs{A}}_3 }
The coefficients $\alpha_1$, $\alpha_2$ and $\alpha_3$ are real. The eigenvalues of $\tilde{\bs{\Omega}}$ are $w_0=0$ and $w_{\pm} = \pm \sqrt{\alpha_3^2 - 4 \alpha_1 \alpha_2}$.
This shows that one of the eigenvalues of $\tilde{\bs{U}} = \exp(\bs{\tilde{\Omega}})$ is always equal to $u_0=1$, { plotted in green in Fig.~\ref{fig:eigs1}}. Since all the involved coefficients are real, the eigenvalues of $\tilde{\bs{\Omega}}$ can either be all real, or one real and two complex conjugates. If we are in the second case, the simultaneous requirements that they are opposite and complex conjugate of each other, implies that they must be purely imaginary. The two conjugate eigenvalues $w_{\pm}$ are thus either $\pm \lambda$ or $\pm i \lambda$, with $\lambda \in \mathbb{R}$. Since the eigenvalues are continuous functions of the parameters, such as $\tau_H$, the only way they can go from $\pm \lambda$ to $\pm i \lambda$ is by becoming $0$, in which case the three eigenvalues {of $\tilde{\bs{\Omega}}$ }are all $0$. In this point all the eigenvalues of $\tilde{\bs{U}}$ {thus} are equal to $1$. 
\begin{figure}[httb!]
\centering
\subfigure{\label{fig:eigsW}
\includegraphics[width=0.48\textwidth]{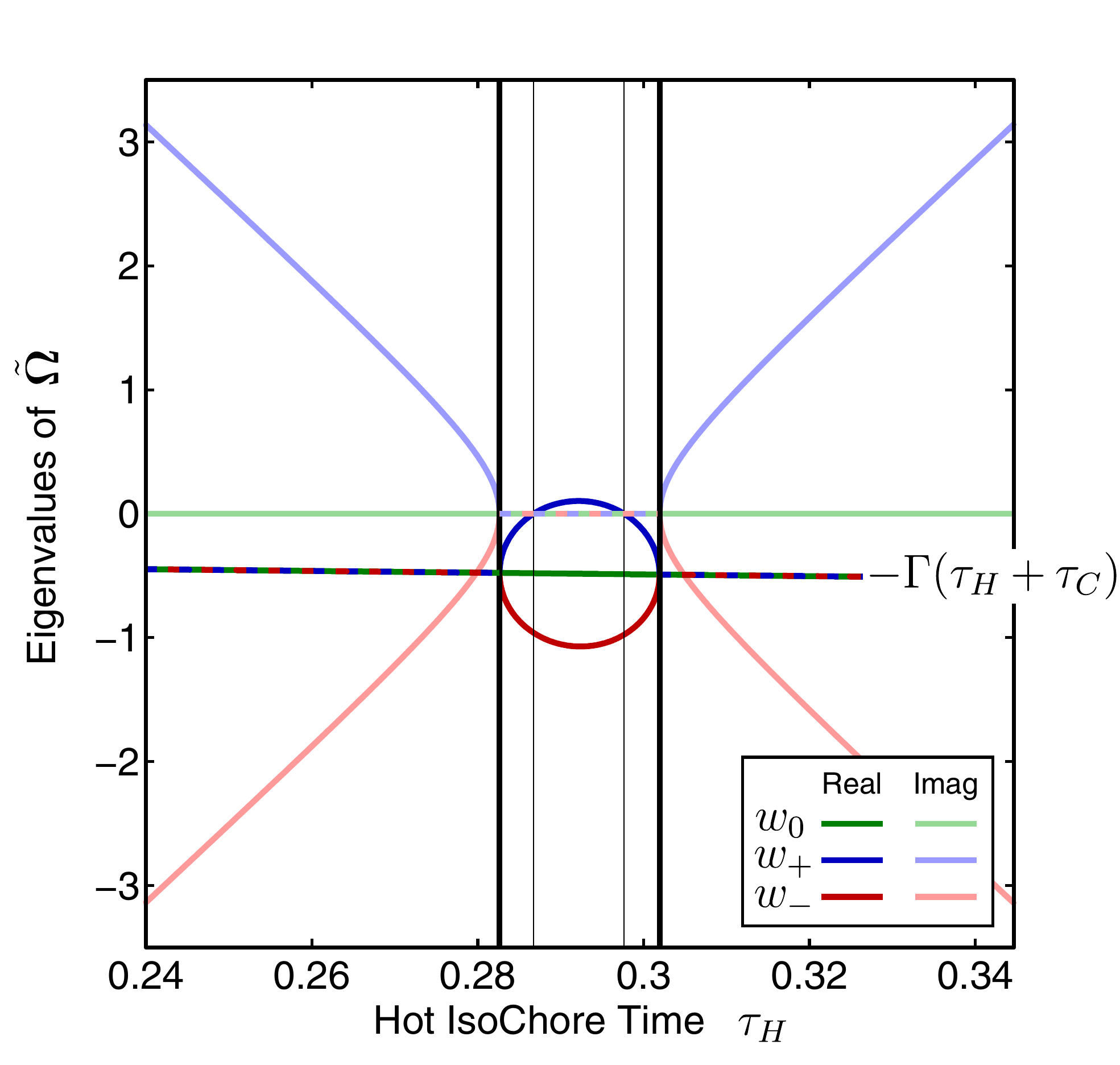}
}
\subfigure{\label{fig:det1t}
\includegraphics[width=0.48\textwidth]{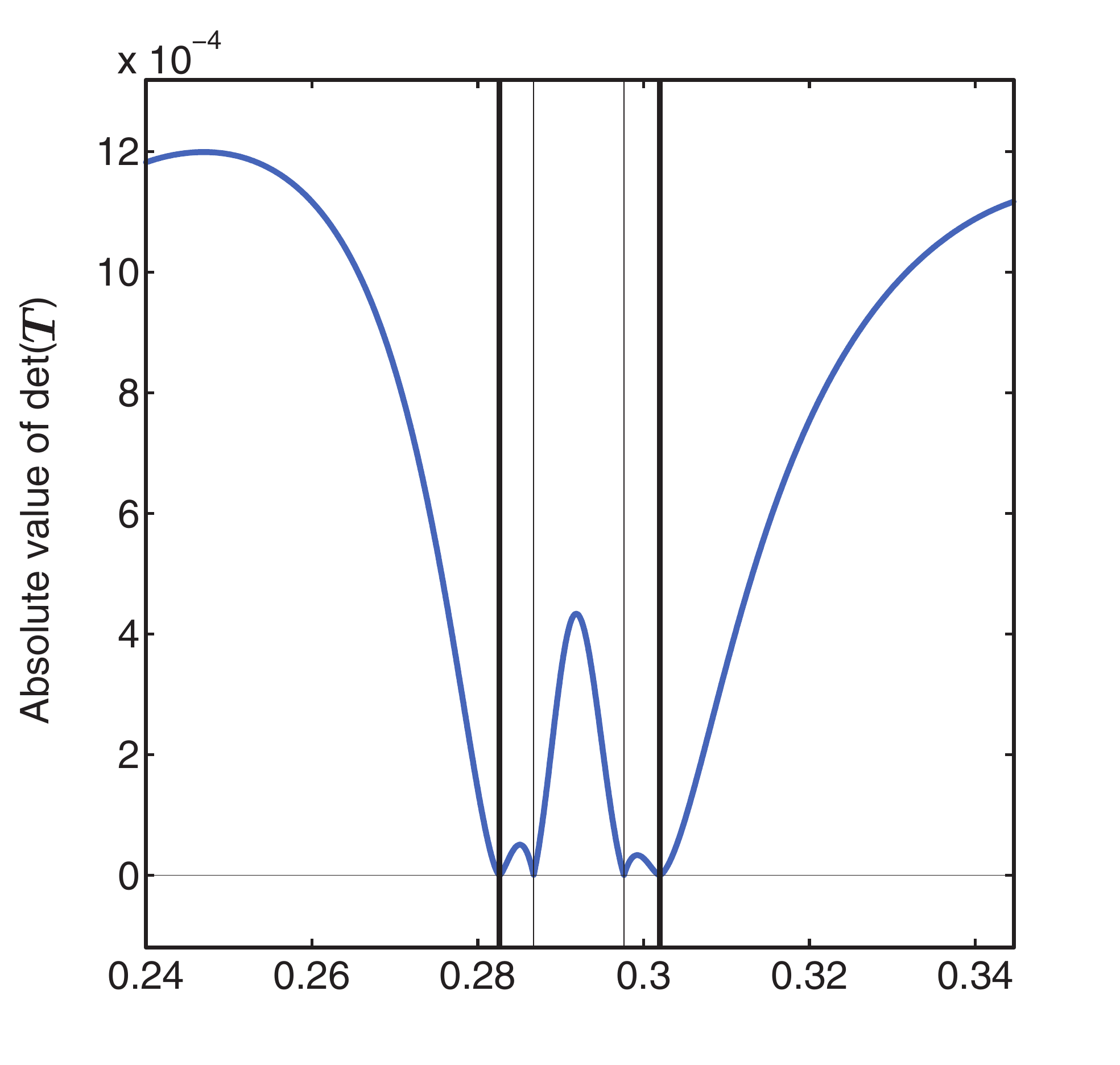}
}
\caption{\emph{Left panel}: eigenvalues of $\tilde{\bs{\Omega}}$. As can be noticed, in the middle region delimited by the thick vertical lines the eigenvalues are purely real. Outside of this region the real part of all the eigenvalues is equal to $-\Gamma(\tau_H+\tau_C)$. At the transition between these two regions all the eigenvalues are exactly equal to $-\Gamma(\tau_H+\tau_C)$. \emph{Right panel}: { the blue curve shows the} absolute value of the determinant $\det(\bs{T})$ of the matrix $\bs{T}$ having as columns the eigenvectors of the time-evolution matrix $\bs{U}(\tau)$. When the determinant is zero we are {at} an exceptional point, i.e. non-hermitian degeneracy. For both panels $\tau_{HC} = \tau_{CH} = 0.1$, and $\tau_C=0.4$ as in Fig.~\ref{fig:eigs1}.} 
\end{figure}

Applying Gaussian elimination on the matrix $\tilde{\bs{\Omega}}(\tau)$, we obtain the following matrix:
\Eq{\tilde{\bs{\Omega}}' (\tau)= \left( \begin{array}{c c c} 
1 &0&-\alpha_1/\alpha_2\\
0&1 &+\alpha_3/\alpha_2\\
0 & 0 & 0 
\end{array} \right)}
Since there are two non-zero rows in $\tilde{\bs{\Omega}}' $, the rank of $\tilde{\bs{\Omega}}$ is always $2$.  The same result remains true as long as at least one of the three coefficients $\alpha_1$, $\alpha_2$, $\alpha_3$ is $\neq 0$.

Because of the rank-nullity theorem, the dimension of the kernel of  $\tilde{\bs{\Omega}}' $  is always $1$. The kernel can also be thought as the eigenspace corresponding to the eigenvalue $0$. A matrix $\tilde{\bs{\Omega}}$ and its exponential $\tilde{\bs{U}}$ always have the same eigenvectors; the eigenvalues of  $\tilde{\bs{U}}$ are the exponential of the eigenvalues of $\bs{\Omega}$. This property is true even in the non-diagonalizable case as it {follows} directly from the definition of the exponential of a matrix: $\tilde{\bs{U}} = \sum_{k=0}^{\infty} \frac{1}{k!}\tilde{\bs{\Omega}}^k$  Therefore the eigenspace associated with the eigenvalue of $\tilde{\bs{U}} $ which is equal to $1$ has dimension always equal to $1$, thus implying three-fold non-hermitian degeneracy  at the transition from complex to real eigenvalues. 

The eigenvalues of $\bs{\tilde{\Omega}}$ are plotted as functions of $\tau_H$ in Fig.~\ref{fig:eigsW}. This figure includes the factor $e^{-\Gamma  ( \tau_H +\tau_C)}$ which has the effect of translating the real part of the eigenvalues of $\bs{\tilde{\Omega}}$ by $-\Gamma  ( \tau_H +\tau_C)$. As can be seen from the figure, the eigenvalue $w_0$, plotted in green, is always equal to $-\Gamma  ( \tau_H +\tau_C)$. Except for this translation, the eigenvalues $w_{\pm}$ are either purely imaginary or purely real, and always opposite of each other. With the translation the real parts are symmetric with respect to the line $-\Gamma  ( \tau_H +\tau_C)$. The transition between real and imaginary is indicated by the thick vertical black lines.

In order to confirm the presence of non-hermitian degeneracy we consider the matrix $\bs{T}$ having the eigenvectors of $\bs{U}$ as columns.  The signature of non-hermitian degeneracy is vanishing of the determinant: when $\bs{U}$ is not diagonalizable, $\bs{T}$ is singular since two or more of its columns are linearly dependent. The absolute value of the determinant of $\bs{T}$ is plotted {in blue} in Fig.~\ref{fig:det1t}. The determinant vanishes for the values of $\tau_H$ indicated by the thick vertical black lines, indicating the transition between real and complex eigenvalues. We already notice that the determinant is also zero on the points indicated by the thin vertical black lines, and this is the subject of the next section.

\subsection{Transition to divergent behaviour: \\$2^{\text{\scriptsize nd}}$ order non-hermitian degeneracy}
In this section we consider the fourth column of the matrix $\bs{U}$, and we will show that \emph{the transition between convergent and divergent behaviour involves an exceptional point.} This transition corresponds to, e.g.,  the values of $\tau_H$ indicated by the thin vertical black lines of Fig.~\ref{fig:eigs1}. In the region where the eigenvalues are real, and at least one of the eigenvalues is larger than $1$, the equilibrium point is unstable.
 
We now consider the full $4\times 4$ matrix $\bs{\Omega}$, still omitting the $e^{-\Gamma(\tau_H+\tau_C)}$ factor for now. When the fourth coordinate is included the most general form of matrix $\bs{\Omega}$  can be written as:
\Eq{ \bs{\Omega} = \left( \begin{array}{c c c c} 
+\alpha_3 &+\alpha_1&0&c_1\\
-2\alpha_2&0 &+2\alpha_1&c_2\\
0&-\alpha_2 &-\alpha_3&c_3\\
0 & 0 & 0&0 
\end{array} \right)}
The eigenvalues of $\tilde{\bs{\Omega}}$ were $w_0=0$ and $w_{\pm}=\pm \sqrt{\alpha_3^2 -4 \alpha_1 \alpha_2 }$. The matrix $\bs{\Omega}$ has one additional eigenvalue which is equal to $0$, (see Sec.~\ref{sec:PropertyBlocks}).  Gaussian elimination gives:
\Eq{ \bs{\Omega}' = \left( \begin{array}{c c c c} 
1 &0&-\alpha_1/\alpha_2&0\\
0&1 &+\alpha_3/\alpha_2&0\\
0 & 0 & 0 &1\\
0 & 0 & 0 &0
\end{array} \right)}
The rank of $\bs{\Omega}$ is thus $3$, and therefore the eigenspace associated with the degenerate eigenvalue $0$ has dimension $1$. {Therefore, whenever two of the eigenvalues of the $4 \times 4$ matrix $\bs{\Omega}$ are simultaneously equal to $0$, a second order non-hermitian degeneracy is present. Since $w_0=0$,} this degeneracy would always be present if it was not for the $e^{-\Gamma (\tau_H+\tau_C)}$ factor multiplying the first three eigenvalues of $\bs{U}$. The quantity $\Gamma  (\tau_H+\tau_C)$ is subtracted from the first three eigenvalues of $\bs{\Omega}$. The degeneracy can thus only appear when $w_{+}$ is equal to $0$, corresponding to the point where $u_{+}$ is $1$. This point is where the transition from convergent to divergent behaviour occurs. As can be seen from Fig.~\ref{fig:det1t}, the determinant vanishes for the values of $\tau_H$ indicated by the thin vertical black lines, indicating the transition from convergent to divergent behaviour.

\section{Existence of limit cycle}
\subsection{Sufficient condition on the structure constant}
We will now show that \emph{when the structure constant is invariant under cyclic permutation of the indices, the existence of a limit cycle is guaranteed.}

The matrix exponential of a skew-symmetric matrix is an orthogonal matrix, and the eigenvalues of an orthogonal matrix always have absolute value equal to $1$. Because of the results of Sec.~\ref{sec:LimitCycle}, we can focus on the matrix $\tilde{\bs{\Omega}}$ appearing in the Magnus expansion. 

Remembering that the commutator between two skew-symmetric matrices is also skew symmetric, we conclude that if  $\tilde{\bs{A}}$ is skew symmetric then all the terms $\bs{\Omega}_k$ appearing in the Magnus expansion are skew symmetric, and so is the sum $\bs{\Omega}$.

In order for the matrix $\bs{A}$ to be skew symmetric, the structure constant $\Gamma_{hj}{}^{k}$ must be anti-symmetric with respect to an exchange between the indices $j$ and $k$.  The structure constant is always anti-symmetric in the first two indices, $\Gamma_{hj}{}^{k} =- \Gamma_{jh}{}^{k}$, since this corresponds to exchanging the operators in the commutator of the left-hand side of its definition, given by Eq.~\ref{eq:GammaDefinition}.  If the structure constant is also invariant under cyclic permutations of the indices, then it is completely anti-symmetric in all indices. In fact, exchanging $j$ and $k$ would give:
\Eq{\Gamma_{hk}{}^{j} = \Gamma_{jh}{}^{k}=-\Gamma_{hj}{}^{k}}
As we will see in Sec.~\ref{sec:SpinSystem}, the structure constant of the spin system satisfies this property and the existence of a limit cycle is guaranteed.

\subsection{Sufficient condition on the Lie algebra}
In this section we discuss the invariance of the structure constant under cyclic permutation of the indices. In particular, we review a sufficient condition for this property to be verified. This condition defines a class of Lie algebras which guarantees the invariance property: \emph{for a compact semisimple Lie algebra there is always a basis for which the structure constant is invariant under cyclic permutation of the indices}.
We assume a finite-dimensional Lie algebra $\mf{g}$ defined over the field of the real numbers $\mathbb{R}$. It is convenient to work with the adjoint representation, whose generic elements will be denoted $X$ and $Y$. The killing form in the adjoint representation is the symmetric bilinear form $K$ defined as:
\Eq{K(X,Y) = \text{Trace}_{\mf{g}}(X  Y)}
The notation $ \text{Trace}_{\mf{g}}$ has the purpose of stressing that the trace is to be intended with respect to the finite-dimensional vector space of the elements composing the Lie algebra $\mf{g}$.
 
Since a representation is a homeomorphism between Lie algebras, the structure constant of the adjoint representation is the same as the one for the original Lie algebra. By Cartan's criterion for semisimplicity, a finite-dimensional real Lie algebra is semisimple if and only if the killing form is non-degenerate \cite{IntroLieAlgebras}. Moreover, it can be shown that the killing form of a compact Lie algebra is negative semi-definite  \cite{WoitNotes}. These two properties together imply that the killing form of a finite-dimensional compact semi-simple real Lie algebra is negative definite.

Since the killing form $K$ is always a symmetric and bilinear form, when it is also definite it can be used it to construct a scalar product. Therefore the scalar product between two elements $X$ and $Y$ can be defined as:
\Eq{\langle X | Y \rangle = - K(X,Y)}
Once the algebra has been equipped with a scalar product, one can choose an orthonormal basis $\{A_k\}$. Such a basis can always be extracted from an arbitrary basis by means of the Gram-Schmidt process. The scalar product between two elements $A_i$ and $A_j$ is thus given by:
\Eq{\langle A_i | A_j \rangle =-K_{ij}= -\text{Trace}(A_i A_j)  = \delta_{ij}\label{eq:TraceDelta}}

We now review the derivation discussed in Ref. \cite{TotalAntiSymm}. We start by considering the commutator between two elements, expressed in terms of the structure constant:
\Eq{[A_j,A_k] = \sum_i \Gamma{}_{jk}{}^{i}\;A_i}
It is then possible to take advantage of the property expressed by Eq.~\ref{eq:TraceDelta}, and write:
\Eq{\text{Trace}(A_l [A_j,A_k] ) =  \sum_i \Gamma{}_{jk}{}^{i}\; \text{Trace}(A_l A_i) = \sum_i \Gamma{}_{jk}{}^{i} (-\delta_{li}) = -\Gamma{}_{jk}{}^{l} \label{eq:CyclicGammaA}}
We can exploit the cyclic property of the trace to manipulate the same expression in a different way:
\Eq{ \setstretch{1.5} \begin{array}{c}
\text{Trace}(A_l [A_j,A_k] ) = \text{Trace}(A_l\; A_j  A_k) - \text{Trace}(A_l \;A_k A_j) =\dots\\
\dots = \text{Trace}( A_k A_l\; A_j ) - \text{Trace}(A_lA_k   \;A_j) = \text{Trace}([A_k,A_l] A_j)  =\dots\\
\dots = \sum_i \Gamma{}_{kl}{}^{i}\;  \text{Trace}(A_i A_j) = \sum_i \Gamma{}_{kl}{}^{i}\; (-\delta_{ij}) =-\Gamma{}_{kl}{}^{j}\end{array} \label{eq:CyclicGammaB}}
Since the starting point of Eq.~\ref{eq:CyclicGammaA} and Eq.~\ref{eq:CyclicGammaB} is the same, we can equate their respective results. Removing the minus sign gives:
\Eq{ \Gamma{}_{jk}{}^{l} = \Gamma{}_{kl}{}^{j} }
which expresses the cyclic property of $\Gamma$. In conclusion, as long as the Lie algebra of operators is finite-dimensional, compact, and semisimple there is a basis under which the structure constant is invariant under cyclic permutation of the indices, and thus completely anti-symmetric.

As a counter-example we consider the harmonic oscillator. As can be calculated from the matrices $\{\bs{A}_h\}$ defined in Eq.~\ref{eq:SetA123}, the matrix {representation} of the killing form for the corresponding algebra is given by:
\Eq{\bs{K} =\hbar^2 \left( \begin{array}{c c c} 
0 &0&-16\\
0&+32 &0\\
-16 & 0 & 0
\end{array} \right)  }
The eigenvalues of $\bs{K}$ are $32\,\hbar^2$ and $\pm\,16\,\hbar^2$, showing that the killing form is indefinite. Therefore, the arguments presented in this section do not apply to the harmonic oscillator.

\subsection{Dimensionality of the Hilbert space}
As we will argue in the present section, \emph{a finite-dimensional Hilbert space does not admit divergent behavior.}
We consider a finite-dimensional Hilbert space $\mc{H}$ over the field $\mathbb{C}$ of the complex numbers. We will argue that the real Lie algebra $\mf{u}(M)$ of all anti-hermitian operators over $\mc{H}$ has dimension $M^2$ and there is a basis for which the structure constant is completely antisymmetric.  Let $M$ be the dimensionality of the Hilbert space and the set $\{| \psi_m \rangle \}_{m=1=,\dots,M}$ be an orthonormal basis. A basis for the real vector space of all anti-hermitian operators is given by:
\Eq{ \setstretch{1.5} \begin{array}{l} 
\hat{X}_{n} =i | \psi_n \rangle \langle \psi_n |, \quad \text{with} \; 1\leq n \leq M\\
\hat{Y}_{nm} = \frac{1}{\sqrt{2}} (| \psi_n \rangle \langle \psi_m |- | \psi_m \rangle \langle \psi_n | ), \quad \text{with} \; 1\leq n < m \leq M\\
\hat{Z}_{nm} = \frac{i}{\sqrt{2}} (| \psi_n \rangle \langle \psi_m |+ | \psi_m \rangle \langle \psi_n | ), \quad \text{with} \;  1\leq n < m \leq M \\
\end{array}}
We thus have the $M$ diagonal operators $\hat{X}_{n}$, the $M(M-1)/2$ ``anti-symmetric'' operators $\hat{Y}_{nm}$, and the $M(M-1)/2$ ``symmetric'' operators $\hat{Z}_{nm}$. All together there are thus $M^2$ anti-hermitian operators which we will collectively denote by $\{A_n\}$. This algebra is the generator of the unitary group $U(M)$, and it can be shown that it is compact. However, the algebra is not semisimple, since it contains the operator $i \hat{1}$ which commutes with all the remaining operators.  This operator forms a one-dimensional abelian ideal of  $\mf{u}(M)$ which prevents the algebra from being semisimple. 

The lack of this property does not constitute an issue: it is possible to extract a set of $M-1$ traceless independent operators $\{\hat{\chi}_{n}\}_{n=1,\dots,M-1}$ from the set $\{\hat{X}_{n}\}_{n=1,\dots,M}$ such that the resulting sub-algebra is compact \emph{and} semisimple. The killing form of this sub-algebra is thus negative definite. The resulting $(M^2-1)$-dimensional algebra $\mf{su}(M)$ is the generator of the special unitary group $SU(M)$.  The most well-known basis is given by the generalized Gell-Mann matrices\cite{BlochQudits,BlochNLevel}: 
\Eq{\hat{\chi}_{n} =\left(\frac{2}{n(n+1)}\right)^{1/2}\Big( -n|\psi_{n+1}\rangle \langle \psi_{n+1} |+ \sum_{k=1}^{n} | \psi_k \rangle \langle \psi_k |\Big), \quad \text{with} \; 1\leq n  \leq M-1}
It can be shown that, over this basis, the structure constant of the algebra is completely anti-symmetric\cite{BlochNLevel}. Since the operator $i\hat{1}$ commutes with any operator, when it is re-introduced in the set of operators the structure constant will not {lose} the property of being completely anti-symmetric.

It can also be shown that any sub-algebra of an algebra whose killing form is negative definite satisfies the same property. We consider again the calculation of the killing form {in} the adjoint representation. If the killing form $K$ is negative definite there is a basis $\{A_j\}_{j=1,\dots,N}$ over which its matrix elements $K_{mn}$ are given by:
\Eq{K_{mn} = \text{Trace} (A_m A_n ) = -\delta_{nm}}
We now consider a rectangular matrix $\bs{C}$ which constructs the sub-algebra $\{A'_j\}_{j=1,\dots,N'<N}$ from the original algebra:
\Eq{A'{}_j = \sum_{m=1}^{N} C_{jm} A_m,\quad \text{with} \quad j =1,\dots, N'< N} 
The matrix elements $K'{}_{jk}$ of the new killing form can be calculated from the following equation: 
\Eq{K'{}_{jk} = \text{Trace} (A'{}_j A'{}_k ) = -\sum_{nm} C_{jm}C_{kn} \delta_{nm}=  - \sum_{n} C_{jn}  C_{kn} } 
In matrix form the killing form of the sub-algebra expanded over the basis $\{A'_j\}_{j=1,\dots,N'<N}$ is thus expressed as:
\Eq{\bs{K}' =- \bs{C} \bs{C}^T}
A matrix of the form $\bs{C} \bs{C}^T$ can be shown to be always symmetric:
\Eq{ (\bs{C} \bs{C}^T)^T = (\bs{C}^{T})^{T} \bs{C}^T= \bs{C} \bs{C}^T}
Moreover, $\bs{C} \bs{C}^T$ is always positive semi-definite:
\Eq{\ul{x}^T \bs{C} \bs{C}^T \ul{x} = (\bs{C}^T \ul{x})^{T} (\bs{C}^T \ul{x}) \geq 0 }
The equality can only occur for a non-zero vector $\ul{x}$ if $\bs{C}$ is singular. If the matrix $\bs{C}$ defines a basis for the subalgebra it must be non-singular, thus guaranteeing that $\bs{C} \bs{C}^T$ is positive definite, and that the killing form $\bs{K}'$ is negative definite.  

It is worth mentioning that the expectation value of any hermitian operator $\hat{L}$ defined over a finite-dimensional Hilbert space $\mc{H}$ has an upper and a lower limit: 
\Eq{\langle \hat{L} \rangle = \sum_{m=1}^{M} p_m L_m}
Here $p_m$ denotes the probability associated with the eigen-ket corresponding to the eigenvalue $L_m$. Denoting by $|L_m\rangle$ the eigen-ket corresponding to the eigenvalue $L_m$, and by $\hat{\rho}$ the density operator, the probability $p_m$ is given by:
\Eq{p_m = \text{Trace} \big(\hat{\rho} | L_m \rangle \langle L_m| \big)}
Since the probabilities satisfy  $0\leq p_m \leq 1$ and $\sum_m p_m=1$, the upper limit of $\langle \hat{L} \rangle$ is given by the largest eigenvalue of $\hat{L}$, and the lower limit is given by its smallest eigenvalue. This argument alone would be sufficient {to exclude the possibility of diverging to infinity.}

One would be tempted to apply the same arguments to infinite-dimensional Hilbert spaces. However, since the trace of an operator defined over an infinite-dimensional space $\mc{H}_{\infty}$ might not exist, it is not guaranteed that the series involved in the previous derivations are convergent. For this reason not all the algebras of anti-hermitian operators over $\mc{H}_{\infty}$ are characterized by a negative-definite killing form.

\subsection{Comparison with the spin system}\label{sec:SpinSystem}
We now consider the case of two coupled spin systems in presence of an external oscillating magnetic field. This system can be treated by considering the following algebra of time-independent hermitian operators \cite{kosloff03}:
\Eq{[\hat{B}_1,\hat{B}_2] =+\sqrt{2} i \hat{B}_3}
\Eq{[\hat{B}_2,\hat{B}_3] =+\sqrt{2} i \hat{B}_1}
\Eq{[\hat{B}_3,\hat{B}_1] =+\sqrt{2} i \hat{B}_2}
It is apparent that the structure constant $\Gamma_{hj}{}^{k}$ is invariant under cyclic permutation of the indices and therefore is completely anti-symmetric. As can be explicitly calculated, the matrix {representation} of the killing form over this basis is proportional to the identity matrix {and is thus negative} definite. 
The Hamiltonian operator governing this system is defined as:
\Eq{\hat{H} = \hbar \omega(t) \hat{B}_1 + \hbar J \hat{B}_2}
The equation of motion can then be written in matrix form as in the following equation:
\Eq{\frac{d}{dt}\left( \begin{array}{c} B_1 \\ B_2 \\ B_3 \end{array} \right) = 
 \left( \begin{array}{c c c} 
0 &0 & +J\\
0&0 &-\omega\\
-J & +\omega & 0 
\end{array} \right) \left( \begin{array}{c}B_1 \\B_2 \\ B_3\end{array} \right)}
Instead of the set of matrices $\bs{A}_1$, $\bs{A}_2$ and $\bs{A}_3$, defined in Eq.~\ref{eq:SetA123}, we see that $\bs{A}$ belongs to the semisimple compact algebra $\mf{so}(3)$ of $3\times 3$ skew-symmetric matrices, which generates the group of rotations SO(3).

As for the harmonic case, the equation of motion which describes the isochoric steps must include the identity operator as fourth element of the algebra. The evolution matrix is modified by subtracting the matrix $\Gamma \bs{A}_0$ defined in Eq.~\ref{eq:A0martix}, and by populating the first three entries of the fourth column with expressions which include $\Gamma$ and the equilibrium energy $H_{\text{eq}}$:
\Eq{\frac{d}{dt}\left( \begin{array}{c} B_1 \\ B_2 \\ B_3\\1 \end{array} \right) = 
 \left( \begin{array}{c c c c} 
-\Gamma &0 & +J& \frac{\Gamma \omega}{\Omega^2}H_{\text{eq}}\\
0&-\Gamma &-\omega& \frac{\Gamma J}{\Omega^2}H_{\text{eq}}\\
-J & +\omega & -\Gamma&0\\
0&0&0&0 
\end{array} \right) \left( \begin{array}{c}B_1 \\B_2 \\ B_3\\ 1 \end{array} \right)}
where the constant $\Omega$ is given by: $\Omega = \sqrt{\omega^2+J^2}$, and the equilibrium energy is $H_{\text{eq}}=\Omega \,\text{tanh}(-\Omega \beta/2)$. The set of parameters used for the calculations of this section are:
\begin{equation}
\label{eq:TheParamSet}
\omega_H = \sqrt{41}, \quad \omega_{C} = \sqrt{11}, \quad \beta_H =0.008, \quad \beta_{C} = 0.03, \quad \Gamma_{H} = \Gamma_{C} = 0.2, \quad J = 2
\end{equation}
The closed form of the limit cycle can be determined exactly in the same way as for the harmonic oscillator. Because of the results of the previous sections we already know that the limit cycle exists for every possible choice of parameters.
\begin{figure}[httb!]
\centering

\subfigure{\label{fig:SpinPow}
\includegraphics[width=0.48\textwidth]{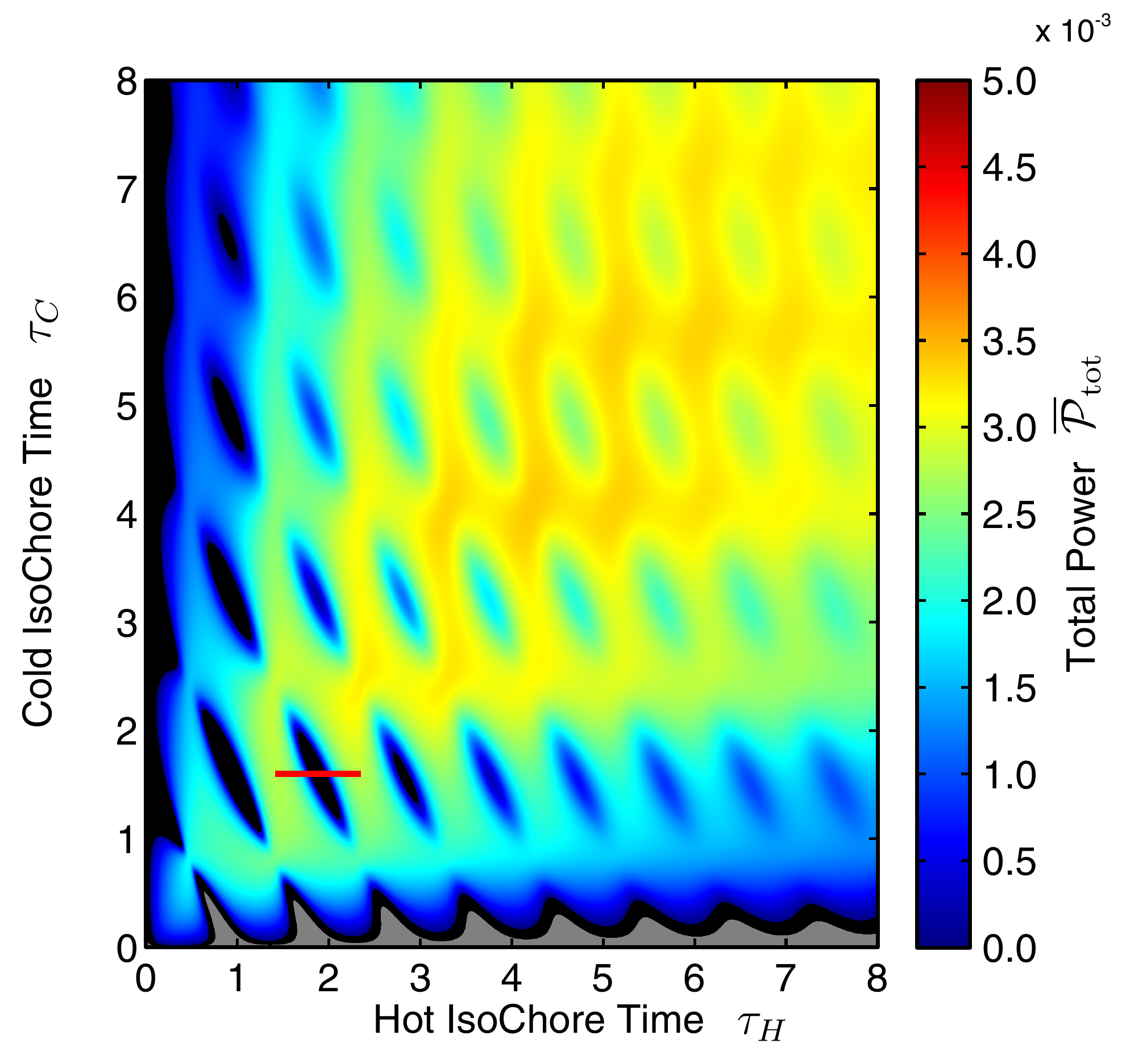}
}
\subfigure{\label{fig:Spineigs1}
\includegraphics[width=0.48\textwidth]{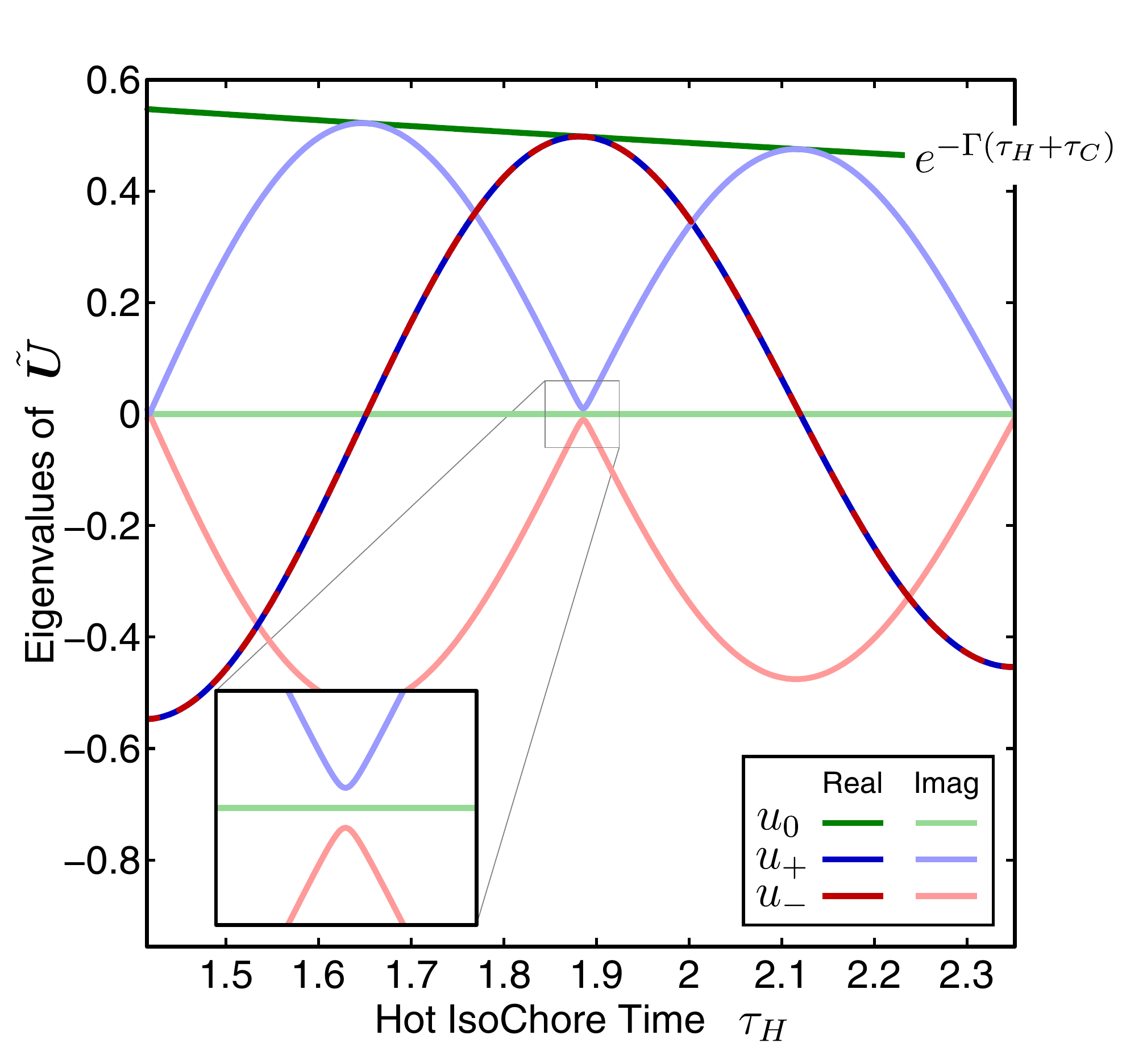}
}

\caption{\emph{Left panel}: power landscape for two coupled spins in presence of an oscillating magnetic field. This  system can never exhibit divergent behavior and {indeed} the white regions visible in Fig.~\ref{fig:Pow} are not present here.  The adiabat times are: $\tau_{HC} = \tau_{CH} = 0.64$. \emph{Right panel}:  eigenvalues of the $3\times 3$ block $\tilde{\bs{U}}$ of the time-evolution matrix for one cycle. For the spin system the moduli of the eigenvalues are always equal to $e^{-\Gamma (\tau_H+\tau_C)}<1$, ensuring the existence of a limit cycle. This panel corresponds to the segment highlighted by the horizontal red line shown in the left panel, i.e.: $\tau_C = 1.6$.} 
\end{figure}

The power landscape for the spin system  as {a} function of the {isochore} times $\tau_H$ and $\tau_C$ is shown in Fig.~\ref{fig:SpinPow}. As can be noticed, the white islands indicating divergent behavior are not present in this case.
The eigenvalues of $\tilde{\bs{U}}$ are plotted as {functions} of $\tau_H$ in Fig.~\ref{fig:Spineigs1}. For the spin system, the moduli of all the eigenvalues are always equal to $e^{-\Gamma (\tau_H+\tau_C)}$. Notice that in the middle point where the eigenvalues $u_{+}$ and $y_{-}$ are \emph{almost} equal, they actually lie on opposite sides of the zero line. Even in the case of a triple degeneracy, $\tilde{\bs{U}}$ could only become proportional to the identity matrix and the degeneracy would be hermitian.

\section{Discussion and conclusions}
The equations of motion of open quantum systems as described by the Lindblad formalism are linear, as they are for closed systems. The study of the limit cycles of quantum heat machines is thus analogous to the classification of equilibrium points of linear dynamical systems. The stability of the equilibrium points is linked to the eigenvalues of the time-evolution matrix for one cycle: as long as all the eigenvalues have modulus smaller than $1$ the equilibrium is stable, but as soon as one of the eigenvalues {has} modulus greater than $1$ we can observe divergent behavior.

From a classical point of {view, it} it is not surprising that a periodically driven dynamical system can be prevented from reaching a steady regime by opportunely selecting the parameters of the {periodic} driving force. {The simplest example is probably the undamped harmonic oscillator sinusoidally driven at its resonance frequency.} Here we observe a singularity in the linear response function which physically means that the induced oscillations will keep increasing in amplitude, without ever reaching a limit-cycle. For the case of a sinusoidal driving force, as long as the damping is not zero, this divergent behavior is not possible: we can always find an equilibrium point between the opposing trends of the damping and driving forces. {More generally,} we can imagine many examples of classical physical systems which, despite the presence of damping, can be driven by a periodic excitation without ever reaching the steady state regime. This happens when the energy dissipation caused by the damping is not enough to counteract the energy pumped into the system by the driving force. As we have shown in the present paper, this behavior is also seen in an ensemble of quantum harmonic oscillators undergoing an Otto cycle.

One of the peculiarities of finite-dimensional quantum systems is the presence of an upper and lower bound to the expectation values of any observable. This is due to the fact that the spectra of the corresponding Hermitian operators, i.e. the possible outcomes of measurements of the observables, are finite sets. Intuitively this implies that it is not possible to observe divergent behavior for such systems. {Employing the formalism of Lie algebras,} we studied the sufficient conditions for a system which cannot exhibit divergence. If the underlying algebra of operators is compact and semisimple, the killing form is negative definite. When this is the case, there is a basis over which the structure constant $\Gamma_{ijk}$ is completely anti-symmetric in all indices, and the corresponding equations of motions will be described by a skew-symmetric matrix $\bs{A}$. Such a matrix always leads to an orthogonal time-evolution matrix $\bs{U}(\tau)$. When such a system is coupled to heat reservoirs providing a source of decoherence, the repeated application of the same thermodynamic cycle will bring it closer and closer to the steady-state regime. This is the case of the spin-system discussed in Sec.~\ref{sec:SpinSystem}.

On the other hand, an infinite-dimensional system is not guaranteed to obey the properties mentioned above. We analysed this aspect of finite-time quantum thermodynamics by studying the most well-known quantum heat machine whose underlying Hilbert space is infinite-dimensional: a heat engine having an ensemble of independent harmonic oscillators as working medium. For some choices of the parameters governing its evolution, here the times allocated for the four steps composing the cycle, the system is unable to reach a steady-state regime. Under these conditions the expectation values of the observables describing the state of the system are unbounded: repeated application of the cycle will lead to larger and larger values. 

The transition from convergent to divergent behavior happens when the modulus of one the eigenvalues of the time-evolution matrix $\bs{U}(\tau)$ becomes larger than one. As we argued in the present work, if we start from a regime where the eigenvalues are complex numbers of modulus smaller than one, before reaching the divergent behavior we encounter a transition to purely real eigenvalues. This transition is characterized by a three-fold non-hermitian degeneracy, i.e. three eigenvalues are equal to $e^{-\Gamma(\tau_H+\tau_C)}$, and the three corresponding eigenvectors simultaneously coalesce. {The coalescence is due to the non-compact algebra and linked to the fact that the Hamiltonian is explicitly time-dependent. This point would in fact be exceptional even without the thermal coupling of the system with the heat reservoirs \cite{k282}}. 

Moreover, the transition to the divergent regime is characterized by {an additional} two-fold non-hermitian degeneracy, when two eigenvalues become equal to $1$ and the corresponding eigenvectors coalesce. {In this case the coalescence is due to the non-hermitian dynamics describing the dissipative interaction of the system with the heat reservoir. As long as thermal coupling is present, this kind of degeneracy can also be observed for quantum systems described by a compact Lie algebra \cite{k301}.}
 
As in previous works on the topic of exceptional points\cite{k282,k301}, the occurrence of non-hermitian degeneracy indicates the transition between two critically different behaviors: {the three-fold non-hermitian degeneracy corresponds to the point where the stationary solution goes from a stable spiral to a stable node; the two-fold non-hermitian degeneracy corresponds to the point where the stationary solution goes from a stable node to an unstable one.}
The { phenomenon of non-hermitian degeneracy} can only be observed {in the presence} of an explicitly time-dependent Hamiltonian\cite{k282} or in the case of open quantum systems\cite{k301}. As highlighted by our study, the analysis of exceptional points potentially leads to interesting {phenomena}.

\section{Acknowledgments}
Ronnie Kosloff acknowledges the Israel Science Foundation.

\bibliographystyle{unsrt}

\section*{Appendices}
\renewcommand{\thesubsection}{\Alph{subsection}}
\subsection{{Some properties of block triangular matrices with a 1 on the diagonal}}\label{sec:PropertyBlocks}
Let us consider a matrix $\bs{A}$ exhibiting the following block structure:
\begin{equation}
\boldsymbol{A} = 
  \left( \begin{array}{c c c |c} 
 & & & \\
 &\tilde{\boldsymbol{A}}& &\tilde{\underline{B}} \\
 & & &  \\
 \hline 0 &0  &0  &0 
\end{array} \right)
\end{equation}
where  $\tilde{\boldsymbol{A}}$ is a $3 \times 3$ matrix block and $\tilde{\underline{B}}$ is a $3 \times 1$ column vector. One of the eigenvalues of $\bs{A}$ is always $0$. The other three eigenvalues coincide with the eigenvalues of $\tilde{\boldsymbol{A}}$. Also the first three components of the corresponding eigenvectors are the same as those of $\tilde{\boldsymbol{A}}$, while the fourth component of these $3$ eigenvector of $\bs{A}$ is $0$. Nothing can be said, in general, about the eigenvectors of  $\bs{A}$ corresponding to the eigenvalue $0$.
If two such matrices $\bs{A}$ and $\bs{A}'$ are multiplied, the result is a matrix $\bs{A}''$ presenting the same structure.
\Eq{\left( \begin{array}{c c c |c} 
 & & & \\
 &\tilde{\boldsymbol{A}}& &\tilde{\underline{B}}\\
 & & &  \\
 \hline 0 &0  &0  &0 
 \end{array} \right)
\left( \begin{array}{c c c |c} 
 & & & \\
 &\tilde{\boldsymbol{A}}'& &\tilde{\underline{B}}'\\
 & & &  \\
 \hline 0 &0  &0  &0 
 \end{array} \right) = \left( \begin{array}{c c c |c} 
 & & & \\
 &\tilde{\boldsymbol{A}}''& &\tilde{\underline{B}}''\\
 & & &  \\
 \hline 0 &0  &0  &0 
 \end{array} \right)}
Where the block $\tilde{\boldsymbol{A}}''$ is the product of the corresponding blocks of the two matrices $\bs{A}$ and $\bs{A}'$:
\Eq{\tilde{\boldsymbol{A}}'' = \tilde{\boldsymbol{A}}\,\tilde{\boldsymbol{A}}' }
We now consider the matrix exponential $\bs{U} = \exp (\bs{A})$ which is always of the form:
\begin{equation}
\boldsymbol{U} = 
 \left( \begin{array}{c c c |c} 
 & & & \\
 &\tilde{\boldsymbol{U}}& &\tilde{\underline{C}}\\
 & & &  \\
 \hline 0 &0  &0  &1 
 \end{array} \right).
\label{eq:TheUMatrix}
\end{equation}
where the matrix block $\tilde{\boldsymbol{U}}$ is independent of $\tilde{\underline{B}}$ and given by:
\Eq{\tilde{\boldsymbol{U}} = \exp \Big(  \tilde{\boldsymbol{A}}\Big) \label{eq:ExpBlock}}
If $\tilde{\underline{B}}$ is zero, then $\tilde{\underline{C}}$ is also zero. One of the eigenvalues of ${\boldsymbol{U}}$ is always $1$ and the other three eigenvalues coincide with those of $\tilde{\boldsymbol{U}}$. As before,  the first three components of the corresponding eigenvectors are the same as those of $\tilde{\boldsymbol{U}}$, while the fourth component of these $3$ eigenvectors of $\bs{U}$ is $0$.

If a matrix such as $\boldsymbol{U}$ is multiplied by a matrix $\boldsymbol{U}'$ exhibiting an analogous structure, the results obeys the following property:
\Eq{\left( \begin{array}{c c c |c} 
 & & & \\
 &\tilde{\boldsymbol{U}}& &\tilde{\underline{C}}\\
 & & &  \\
 \hline 0 &0  &0  &1 
 \end{array} \right)
\left( \begin{array}{c c c |c} 
 & & & \\
 &\tilde{\boldsymbol{U}}'& &\tilde{\underline{C}}'\\
 & & &  \\
 \hline 0 &0  &0  &1 
 \end{array} \right) = \left( \begin{array}{c c c |c} 
 & & & \\
 &\tilde{\boldsymbol{U}}''& &\tilde{\underline{C}}''\\
 & & &  \\
 \hline 0 &0  &0  &1 
 \end{array} \right)}
Again, the matrix block $\tilde{\boldsymbol{U}}''$ is  independent of $\tilde{\underline{C}}$ and $\tilde{\underline{C}}'$ and is given by the product:
\Eq{\tilde{\boldsymbol{U}}'' = \tilde{\boldsymbol{U}} \, \tilde{\boldsymbol{U}}'\label{eq:ProductBlocks}}

\end{document}